\documentclass[camera,letterpaper,nomarginnotes,nonarrowgutter]{jpaper}

\usepackage{dblfloatfix}
\usepackage{multirow}
\usepackage{booktabs}
\usepackage{array}
\usepackage{cite}
\usepackage{algorithm}
\usepackage{eucal}
\usepackage{xcolor,colortbl}
\definecolor{freakishgreen}{HTML}{0A982B}
\definecolor{urlblue}{HTML}{319dd6}
\usepackage{enumitem}
\usepackage{fancyhdr}
\usepackage{enumitem}
\usepackage[noend]{algpseudocode}
\usepackage{datetime}
\newcolumntype{L}[1]{>{\raggedright\let\newline\\\arraybackslash\hspace{0pt}}m{#1}}
\newcolumntype{C}[1]{>{\centering\let\newline\\\arraybackslash\hspace{0pt}}m{#1}}
\newcolumntype{R}[1]{>{\raggedleft\let\newline\\\arraybackslash\hspace{0pt}}m{#1}}
\usepackage{tikz}
\newcommand*\circled[1]{\tikz[baseline=(char.base)]{
            \node[shape=circle,fill,inner sep=1pt] (char) {\textcolor{white}{#1}};}}
\newcommand*\bluecircled[1]{\tikz[baseline=(char.base)]{
            \node[shape=circle,fill,color=blue,inner sep=1pt] (char) {\textcolor{white}{#1}};}}
\newcommand*\redcircled[1]{\tikz[baseline=(char.base)]{
            \node[shape=circle,fill,color=red,inner sep=1pt] (char) {\textcolor{white}{#1}};}}
\makeatletter
\def\thickhline{%
  \noalign{\ifnum0=`}\fi\hrule \@height \thickarrayrulewidth \futurelet
   \reserved@a\@xthickhline}
\def\@xthickhline{\ifx\reserved@a\thickhline
               \vskip\doublerulesep
               \vskip-\thickarrayrulewidth
             \fi
      \ifnum0=`{\fi}}
\makeatother
\newlength{\thickarrayrulewidth}
\usepackage{pifont}

\newcommand\Tstrut{\rule{0pt}{2ex}}         
\newcommand\Bstrut{\rule[-1ex]{0pt}{0pt}}   
\def\thickhline{\noalign{\hrule height.8pt}}
\newcolumntype{?}{!{\vrule width 0.8pt}}

\usepackage{setspace}
\usepackage{soul}
\definecolor{schrift}{RGB}{0,73,174}

\usepackage{afterpage}

\usepackage{xhfill}

\setul{0.5ex}{0.3ex}
\setulcolor{blue}

\newif\ifsubmission
\submissiontrue

\newif\ifarxiv
\arxivtrue

\ifsubmission
    \newcommand{\rbc}[1]{{#1}}
    \newcommand{\rbd}[1]{{#1}}
    \newcommand{\rbe}[1]{{#1}}
    \newcommand{\rbf}[1]{{#1}}

    \fancypagestyle{firstpage}
    {
        
        \fancyhf{}
        \fancyfoot{}
        \fancyfoot[C]{\thepage}
    }
\else
    \newcommand{\rbc}[1]{{\color{blue}#1}}
    \newcommand{\rbd}[1]{{\color{magenta}#1}}
    \newcommand{\rbe}[1]{{\color{brown}#1}}
    \newcommand{\rbf}[1]{{\color{Green}#1}}

    \newcommand{\versionnum}[0]{4.1}
    \fancypagestyle{firstpage}
    {
        \fancyhead[C]{\textcolor{blue}{\emph{Version \versionnum~---~\today, \xxivtime \ UTC}}}
        \fancyfoot[C]{\thepage}
    }
\fi

\usepackage{hyperref}

\hypersetup{
    bookmarks=true,
    breaklinks=true,
    colorlinks=true,
    linkcolor=blue,
    citecolor=blue,
    urlcolor=urlblue,
    pdfpagelabels=false,
 }

\usepackage[nameinlink]{cleveref}
\crefformat{section}{\S#2#1#3} 
\crefformat{subsection}{\S#2#1#3}
\crefformat{subsubsection}{\S#2#1#3}

\def\BibTeX{{\rm B\kern-.05em{\sc i\kern-.025em b}\kern-.08em
    T\kern-.1667em\lower.7ex\hbox{E}\kern-.125emX}}

\usepackage[symbol]{footmisc}
\renewcommand{\thefootnote}{\fnsymbol{footnote}}

\title{{Constable: Improving Performance and Power Efficiency \\ by Safely Eliminating Load \rbd{Instruction} Execution}}
\ifarxiv
    \author{
    *Rahul Bera$^1$ \hspace{0.5em} *Adithya Ranganathan$^2$ \hspace{0.5em} Joydeep Rakshit$^2$ \hspace{0.5em} Sujit Mahto$^2$ \vspace{0.2em} \\
    Anant V. Nori$^2$ \hspace{0.5em} Jayesh Gaur$^2$ \hspace{0.5em} Ataberk Olgun$^1$ \hspace{0.5em} Konstantinos Kanellopoulos$^1$ \vspace{0.2em} \\
    Mohammad Sadrosadati$^1$ \hspace{0.5em} Sreenivas Subramoney$^2$ \hspace{0.5em} Onur Mutlu$^1$ \vspace{0.5em} \\
    \normalsize{
        $^1$ETH Zürich \hspace{0.5em} $^2$Intel Processor Architecture Research Lab
    }
    }
\else
    \author{
    *$^\dagger$Rahul Bera$^1$ \hspace{0.5em} *Adithya Ranganathan$^2$ \hspace{0.5em} Joydeep Rakshit$^2$ \hspace{0.5em} Sujit Mahto$^2$ \vspace{0.2em} \\
    Anant V. Nori$^2$ \hspace{0.5em} Jayesh Gaur$^2$ \hspace{0.5em} Ataberk Olgun$^1$ \hspace{0.5em} Konstantinos Kanellopoulos$^1$ \vspace{0.2em} \\
    Mohammad Sadrosadati$^1$ \hspace{0.5em} Sreenivas Subramoney$^2$ \hspace{0.5em} Onur Mutlu$^1$ \vspace{0.5em} \\
    \normalsize{
        $^1$ETH Zürich \hspace{0.5em} $^2$Intel Processor Architecture Research Lab
    }
    }
\fi

\begin{document}



\maketitle

\ifarxiv
    \def\thefootnote{}\footnotetext{*\rbf{Rahul Bera and Adithya Ranganathan are co-primary authors}.
    }\def\thefootnote{\arabic{footnote}}
\else
    \def\thefootnote{}\footnotetext{
    *\rbf{Co-primary authors. $^\dagger$This work was partially done when Rahul Bera was an intern at Intel Processor Architecture Research Lab. Concepts, techniques, and implementations presented in this paper may be subject matter of pending patent applications filed by Intel Corporation}.
    }\def\thefootnote{\arabic{footnote}}
\fi

\thispagestyle{firstpage}

\begin{abstract}
Load instructions often limit instruction-level parallelism (ILP) in modern processors due to data and resource dependences they cause. 
Prior techniques like Load Value Prediction (LVP) and Memory Renaming (MRN) mitigate load data dependence by predicting the data value of a load instruction. 
However, they fail to mitigate load resource dependence as the predicted load instruction gets executed nonetheless \rbd{(even on a correct prediction)}, which consumes hard-to-scale pipeline resources that otherwise could have been used to execute other load instructions.

Our goal in this work is to improve ILP by mitigating \emph{both} load data dependence and resource dependence. 
To this end, we propose a purely-microarchitectural technique \rbd{called} \emph{Constable}, that safely eliminates the execution of load \rbd{instructions}. 
Constable dynamically identifies load instructions that have repeatedly fetched the same data from the same load address. We call such loads \emph{likely-stable}.
For every likely-stable load, Constable (1) tracks modifications to its source architectural registers and memory location via lightweight \rbd{hardware} structures, and (2) eliminates \rbc{the execution of} subsequent instances of \rbc{the} load instruction until there is a write to its source register or a store or snoop request to its load address.

Our extensive evaluation using a wide variety of 90 workloads shows that Constable improves performance by $5.1\%$ while \emph{reducing} the core dynamic power consumption by $3.4\%$ on average over a strong baseline system that implements MRN and other dynamic instruction optimizations (e.g., move and zero elimination, constant and branch folding). 
\rbc{In presence of 2-way simultaneous multithreading (SMT), Constable's performance improvement increases to $8.8\%$ over the baseline system.}
When combined with a state-of-the-art load value predictor (EVES), Constable provides an additional $3.7\%$ and $7.8\%$ average performance benefit over \rbc{the load value predictor} alone, in the baseline system without and with 2-way SMT, respectively.



\end{abstract}
\section{Introduction} \label{sec:introduction}

Extracting high instruction-level parallelism (ILP)~\cite{ilp,ilp2} is essential for designing high-performance processors. However, ILP often gets limited by data dependence (i.e., the result of one instruction would be consumed by the other) and resource dependence (i.e., two or more instructions \rbd{contending} for the same limited hardware resource) between instructions~\cite{tjaden1970detection,patt1985hps,jouppi1989available,smith1989limits}.
Load instructions are \rbd{a major} source of ILP limitation in modern workloads due to both data dependence and resource dependence~\cite{austin1995zero}.
On \rbc{the} one hand, a load instruction typically takes longer latency to execute than most non-memory instructions since \rbc{a load performs} two component operations: (1) \rbc{compute} the load address, and (2) \rbc{fetch} the data by accessing the memory hierarchy. 
As a result, the \rbc{instructions that depend on the} load instruction often stall for multiple cycles, \rbc{which can limit} ILP.
On the other hand, a load instruction consumes \rbc{multiple} hard-to-scale hardware resources (e.g., \rbc{reservation station entry}, port to address generation unit, L1-data cache \rbc{read port}) which often causes resource dependence in the pipeline, also limiting ILP. 

Prior works \rbc{propose} many latency tolerance techniques to improve ILP by mitigating load data dependence.
Load Value Prediction (LVP)~\cite{lipasti1996vp,sazeides_vp,sazeides_vp2,mendelson1997speculative,dfcm,last_n,selective_vp,perais2012revisiting,perais2014eole,perais2014practical,perais2015bebop,fvp,perais2021ssr,eves,rami_ap,rami_composite,kalaitzidis2019value,sakhuja2019combining} and Memory Renaming (MRN)~\cite{mrn,mrn2,mrn_classifying,moshovos1997streamlining,moshovos1999speculative} are two such widely-studied techniques that mitigate load data dependence via data value speculation.
\rbc{LVP and MRN speculatively execute load-data-dependent instructions using the predicted value of the load instruction, thus improving ILP.}

\vspace{2pt}
\noindent \textbf{Key limitation.} 
Even though LVP and MRN provide performance benefits by breaking load data dependence, the predicted load instruction \rbc{still} needs to \rbc{be} executed nonetheless to verify the speculated load value, which takes hard-to-scale hardware resources that otherwise could have been utilized for executing other load instructions. In other words, LVP and MRN provide performance benefits by mitigating load data dependence, but they \emph{do not} mitigate load resource dependence, which leaves a significant performance and power \rbc{consumption} improvement opportunity on the table \rbd{(see~\cref{sec:headroom_perf_headroom})}.

\textbf{Our goal} in this work is to improve ILP by mitigating \emph{both} load data dependence and resource dependence. 
To this end, we propose a \rbc{lightweight}, purely-microarchitectural technique \rbc{called} \textbf{\emph{Constable}}, which safely eliminates \rbc{the entire execution of a load instruction} (i.e., \rbc{both} load address \rbc{computation} and \rbc{data fetch from memory hierarchy}).

The \textbf{key insight} behind Constable is that a dynamic load instance $I_2$ of a static load instruction $I$ is bound to fetch the same value from the same memory location as the previous dynamic instance $I_1$ of the same static load instruction when the following two conditions are satisfied.
\begin{itemize}
    \item \textbf{Condition~1}: None of the source registers of $I$ has been written between the occurrences of $I_1$ and $I_2$. 
    \item \textbf{Condition~2}: No store or snoop request has arrived to the memory address of $I_1$ between the occurrences of $I_1$ and $I_2$. 
\end{itemize}

Satisfying Condition~1 ensures that $I_2$ would have the same load address as $I_1$, and thus the address computation operation of $I_2$ can be safely eliminated.
Satisfying Condition~2 ensures that $I_2$ would fetch the same value from the memory as $I_1$, and thus the data \rbc{fetch operation} of $I_2$ can be safely eliminated. 

\vspace{2pt}
\noindent \textbf{Key mechanism.}
Constable exploits \rbd{this key} insight to safely eliminate executing a load instruction while breaking load data dependence in two key steps.
First, Constable dynamically identifies load instructions that have repeatedly fetched the same \rbd{data} value from the same load address. We call such loads \emph{likely-stable}.\footnote{\rbc{Hence the name Constable, that \emph{polices} the likely-stable loads to safely eliminate them}.}
Second, \rbc{when Constable gains enough confidence that a given load instruction is likely-stable,}
Constable tracks modifications to the source architectural registers \rbc{the load instruction} and its memory location via two small \rbd{hardware} structures.
Constable \rbc{eliminates} the execution of \rbc{all future} instances of the likely-stable load and breaks the load data dependence using the last-fetched value \rbc{of the load instruction}, until there is a write to its source registers or a store or snoop request to its load address.

\vspace{2pt}
\noindent \textbf{Key results.}
We evaluate Constable using a diverse set of \emph{90} workloads (which includes all workloads from SPEC CPU 2017~\cite{spec2017} suite and many well-known \texttt{Client}, \texttt{Enterprise}, and \texttt{Server} workloads) over a strong $6$-wide \rbc{superscalar} baseline processor that already implements Memory Renaming and various \rbc{other} dynamic instruction optimizations like zero elimination~\cite{trace_cache}, move elimination~\cite{trace_cache,cont_opt}, constant folding~\cite{trace_cache,cont_opt}, and branch folding~\cite{branch_folding}. Our evaluation yields five key results that show Constable's effectiveness.
First, Constable improves performance on average by (up to) $5.1\%$ ($31.2\%$) over the baseline, while reducing the core dynamic power consumption by $3.4\%$ \rbd{($24.6\%$)}.
Second, when combined with a state-of-the-art value predictor (EVES~\cite{eves}), Constable improves performance by $8.5\%$ on average over the baseline, which is $3.7\%$ more performance than EVES alone. 
Third, in a $2$-way simultaneous multithreading~\cite{smt} configuration, 
\rbc{(a) Constable alone improves performance on average by $8.8\%$ over the baseline, and}
(b) Constable combined with EVES outperforms EVES alone by $7.8\%$ on average.
Fourth, by eliminating both address computation and data \rbc{fetch} component of load execution, Constable reduces the reservation station \rbc{allocations} and L1-data cache \rbc{accesses} by $8.8\%$ and $26\%$, respectively, which results in dynamic power savings.
Fifth, Constable's \rbc{benefits} come at a modest storage overhead of only $12.4$~KB \rbd{per core}.

\vspace{2pt}
\noindent \rbc{We} make the following \textbf{key contributions} in this work:
\begin{itemize}
    \item We show that a significant fraction (\rbd{on average by} $34.2\%$ \rbd{and up to $68.3\%$}) of all \rbc{dynamic} loads repeatedly \emph{\rbc{fetch} the same value from the same memory address across the entire workload}. 
    By analyzing the disassembly of fully-optimized workload binaries, we show that even a state-of-the-art compiler with full optimization often fails to optimize such load instructions at compile time due to various empirically-observed reasons, e.g., accessing runtime constants and accessing local variables in inlined functions (see \cref{sec:headroom_stable_loads_why}).
    \item We show that \rbc{\emph{ideally}} eliminating the execution of such load instructions while breaking their load data dependence has \rbc{more than twice the} performance headroom ($9.1\%$ on average) than breaking \rbd{only} load data dependence via \rbc{\emph{ideal}} load value prediction ($4.3\%$ on average). 
    This performance gap \rbc{demonstrates the opportunity to improve} ILP by mitigating load resource dependence (see \cref{sec:headroom_perf_headroom}).
    \item \rbc{We propose Constable, a microarchitectural technique that dynamically identifies load instructions that repeatedly fetch the same data from the same address using a confidence-based mechanism and eliminates \rbd{the execution of} such instructions by tracking modifications to their source registers and store and snoop requests to their addresses (see \cref{sec:key_idea}).}
    \item We propose a practical, lightweight microarchitecture for Constable that safely eliminates load execution in \rbd{the} presence of aggressive out-of-order load issue in modern multi-core processors (see \cref{sec:design}). 
    \item We extensively verify the correctness of Constable by matching the outcome of every instruction in microarchitectural simulation with that in functional simulation over a large suite of $3400$ traces \rbc{(see \cref{sec:methodology_func_veri})}.
    \item \rbc{We show that Constable improves performance by $5.1\%$ on average (see~\cref{sec:eval_perf_no_smt}) while reducing the core dynamic power consumption by $3.4\%$ (see~\cref{sec:eval_power}) and incurring a modest storage overhead of only $12.4$~KB over a strong superscalar processor that implements various dynamic instruction optimizations.} \rbd{Constable's benefits further increase in presence of simultaneous multithreading (see~\cref{sec:eval_perf_smt}).}
    \item \rbc{We open-source a binary instrumentation tool that we use to identify load instructions that repeatedly fetch the same value from the same memory address in any off-the-self x86-64 binary  in \url{https://github.com/CMU-SAFARI/Load-Inspector} (see~\cref{sec:headroom_workload_analysis})}. 
\end{itemize}
\section{Background}

Extracting high instruction-level parallelism (ILP)~\cite{ilp,ilp2} is essential in providing high single-thread and multi-thread performance in modern processors~\cite{hill2008amdahl,suleman2009accelerating,joao2012bottleneck}. Unfortunately, ILP often gets limited by \emph{data dependence} and \emph{resource dependence} between instructions~\cite{tjaden1970detection,patt1985hps,jouppi1989available,smith1989limits,austin1995zero}.\footnote{
ILP also gets limited by frequent \emph{control dependence}~\cite{tjaden1970detection,patt1985hps,patt1985critical},
which is outside the scope of this work.
} 
Data dependence limits ILP due to data \rbd{flow (communications)} between instructions, whereas resource dependence (also called structural \rbd{dependence}) limits ILP due to \rbd{contention for} limited hardware resources in the system (e.g., execution unit, load port).

\begin{figure}[!h]
\centering
\includegraphics[width=3.4in]{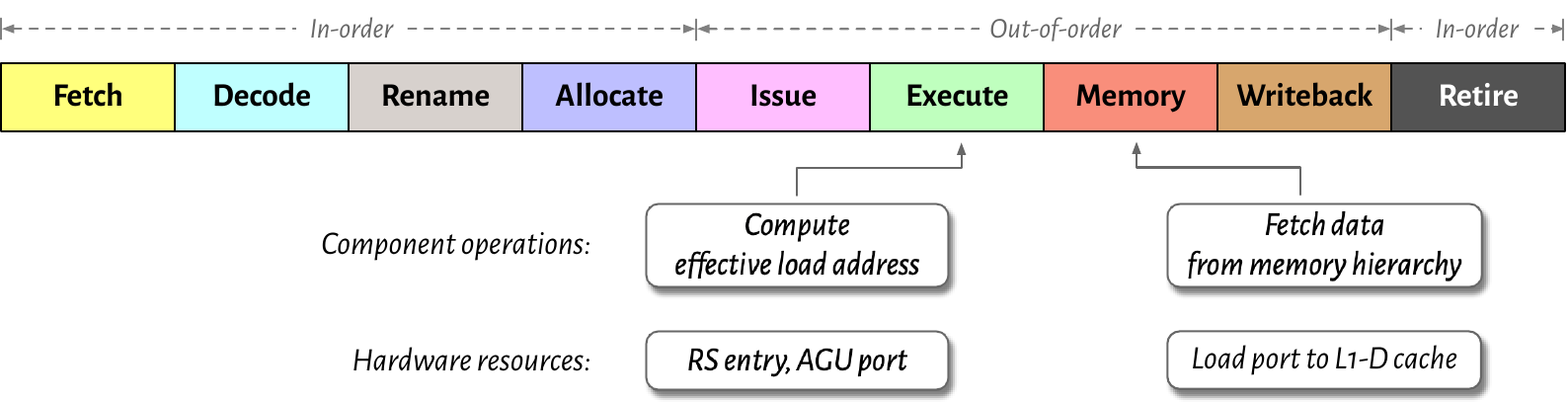}
\caption{
Two component operations of a load instruction execution and their associated pipeline resources. 
}
\label{fig:pipeline}
\end{figure}

Load instructions are \rbd{a major} source of ILP limitation in modern workloads due to both data and resource dependence~\cite{austin1995zero}. 
Load instructions typically have longer latency than most non-memory instructions since they perform multiple component operations in a single instruction. 
Fig.~\ref{fig:pipeline} shows the two key component operations of a load \rbd{instruction's} execution in a traditional nine-stage out-of-order (OOO) processor pipeline~\cite{patt1985hps,hennessy2011computer}. 
The processor first computes the load address of an issued load instruction in the execute stage of the pipeline. The processor then fetches the data by accessing the memory hierarchy in the memory stage, which may require accessing the lower levels of memory hierarchy (e.g., the main memory). As a result, the dependents of a load instruction often stall for several cycles, thus significantly limiting ILP. 

\rbd{A} load \rbd{instruction} also \rbd{uses} several key hardware resources during its execution. As shown in Fig.~\ref{fig:pipeline}, a load instruction consumes (1) a reservation station (RS) entry and an address generation unit (AGU) port to compute its load address, and (2) a load port to the L1-data (L1-D) cache to fetch the data from the memory hierarchy. Many of these hardware resources are hard to scale due to their non-trivial area and power \rbd{overheads}. As a result, a load instruction often causes resource dependence in the pipeline, thus limiting ILP.

Prior works \rbd{propose} many techniques to mitigate load data dependence by tolerating load instruction latency. \emph{Load Value Prediction} and \emph{Memory Renaming} are two such key techniques that mitigate load data dependence via speculative execution. 


\textbf{Load Value Prediction (LVP)}~\cite{lipasti1996vp,sazeides_vp,sazeides_vp2,mendelson1997speculative,dfcm,last_n,selective_vp,perais2012revisiting,perais2014eole,perais2014practical,perais2015bebop,fvp,perais2021ssr,eves,rami_ap,rami_composite,kalaitzidis2019value,sakhuja2019combining} breaks load data dependence by predicting the value of a load instruction and speculatively executing load-data-dependent instructions with the predicted value. The predicted value is later verified by executing the load instruction. A correct prediction increases ILP, but an incorrect prediction leads to re-execution of load-dependent instructions, incurring both performance and power overhead.

\textbf{Memory Renaming (MRN)}~\cite{mrn,mrn2,mrn_classifying,moshovos1997streamlining,moshovos1999speculative} learns \rbd{the} dependence \rbd{relationship} between a store-load instruction pair, and speculatively executes the load-dependent instructions by forwarding the data directly from the associated store instruction. 
\rbd{The forwarded data is later verified by executing the load. A correct data forwarding increases ILP, whereas an incorrect forwarding incurs both performance and power overhead due to re-execution of load-dependent instructions.}
\section{Motivation and Goal}


Even though LVP and MRN provide performance benefit by breaking load data dependence, the predicted load gets executed nonetheless to verify the speculated load value, which takes scarce and hard-to-scale hardware resources that otherwise could have been utilized for executing other load instructions. In other words, LVP and MRN provide performance benefits by mitigating load data dependence, but they \emph{do not} mitigate load resource dependence.

\begin{figure}[!h]
\centering
\includegraphics[width=3.4in]{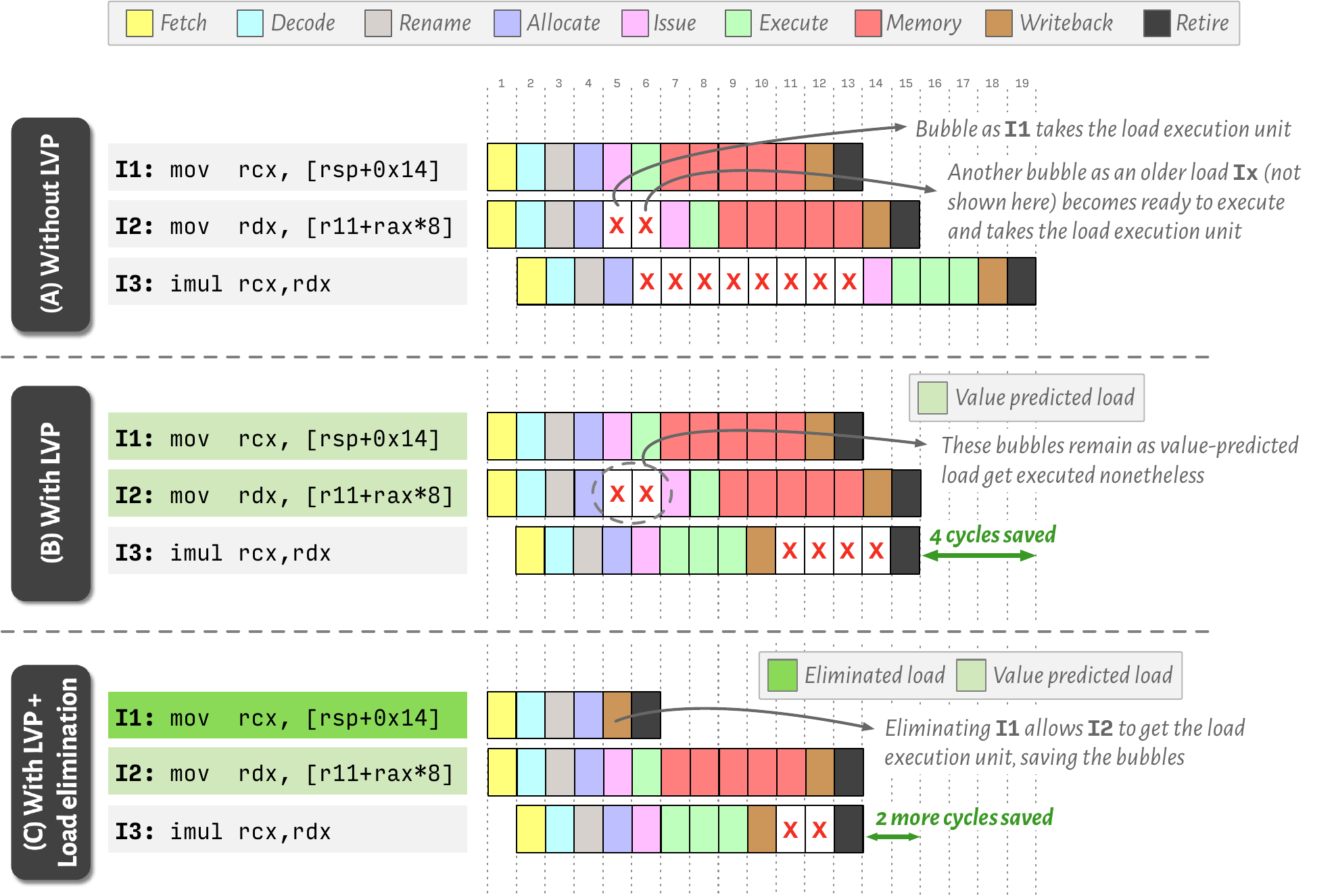}
\caption{
Execution timeline of a code \rbd{example} in a processor (a) without a load value predictor (LVP), (b) with LVP, and (c) with LVP and load elimination.
}
\label{fig:elim_timeline}
\end{figure}

To illustrate how LVP provides performance benefit by mitigating data dependence,\footnote{Since LVP and MRN \rbd{work} on conceptually similar principles, we use LVP for this discussion without loss of generality.} yet the benefit may get limited by resource dependence, Fig.~\ref{fig:elim_timeline}(a) and (b) show the execution timeline of a \rbd{code example} in a processor without and with LVP, respectively. 
For simplicity, we assume that the OOO processor has fetch, issue, and retire bandwidth of two instructions, and one load execution unit (comprised of an AGU and a load port).
We also assume a perfect LVP.
As Fig.~\ref{fig:elim_timeline}(a) shows, $I_1$ gets issued to the load execution unit in cycle-$5$, thus stalling $I_2$. 
In cycle-$6$, an older load instruction $I_x$ (not shown in the figure) becomes ready to execute and gets issued to the load execution unit, thus stalling $I_2$ even further.
Stalls like these, where a load instruction gets delayed due to limited hardware resources (i.e., resource dependence), frequently occur in \rbd{a} modern high-performance processor with deeper and wider pipeline, as we quantitatively show in \cref{sec:headroom_resoruce_hazard}. 
These stalls get exacerbated further in \rbd{the} presence of performance-enhancement techniques like simultaneous multithreading (SMT)~\cite{smt}, where a hardware resource may get shared across SMT threads.

\rbd{When} LVP \rbd{is employed}, as shown in Fig.~\ref{fig:elim_timeline}(b), both loads $I_1$ and $I_2$ get value-predicted and the data-dependent instruction $I_3$ retires 4 cycles earlier than \rbd{in} the processor without LVP.
However, since both $I_1$ and $I_2$ need to get executed to verify their respective predicted values, $I_2$ still experiences stalls in cycle-$5$ and $6$ due to resource dependence.
If we can safely eliminate \rbd{the execution of} $I_1$ while breaking its data dependence, as shown in Fig.~\ref{fig:elim_timeline}(c), \rbd{we can enable} $I_2$ to get issued to the load execution unit in cycle-$5$, which provides an additional $2$ cycles savings on top of the processor with LVP.\footnote{\rbd{Similarly, $I_2$ can potentially be eliminated as well, providing further savings in execution time (not shown in Fig.~\ref{fig:elim_timeline})}.}



\rbd{We conclude} that LVP and MRN may improve performance by mitigating load data dependence, but they leave performance improvement opportunity by not mitigating load resource dependence.

\subsection{Our Goal}
\textbf{Our goal} in this work is to improve ILP by mitigating \emph{both} load data dependence and resource dependence. 
To this end, we propose a \rbc{lightweight}, purely-microarchitectural technique \rbc{called} \textbf{\emph{Constable}}, which safely eliminates \rbc{the entire execution of a load instruction} (i.e., \rbc{both} load address \rbc{computation} and \rbc{data fetch from memory hierarchy}).
\section{Performance Headroom of Constable} \label{sec:headroom}

To understand the performance headroom of Constable, we first study the static load instructions that repeatedly fetch the same value from the same load address \emph{across the entire workload trace}. 
We call such a load \emph{\rbe{global-stable}}. 
Essentially, a \rbe{global-stable} load is a prime candidate for elimination since both load address computation and \rbc{data fetch} operations of its execution produce \rbd{the same} result across all dynamic instances of the instruction. 
We then quantify the resource dependence on \rbe{global-stable} loads \rbd{(\cref{sec:headroom_resoruce_hazard})},
\rbd{and} the performance benefit of \rbc{ideally} eliminating all \rbe{global-stable} load execution \rbd{(\cref{sec:headroom_perf_headroom})}.

\begin{figure*}[!h]
\centering
\includegraphics[width=7in]{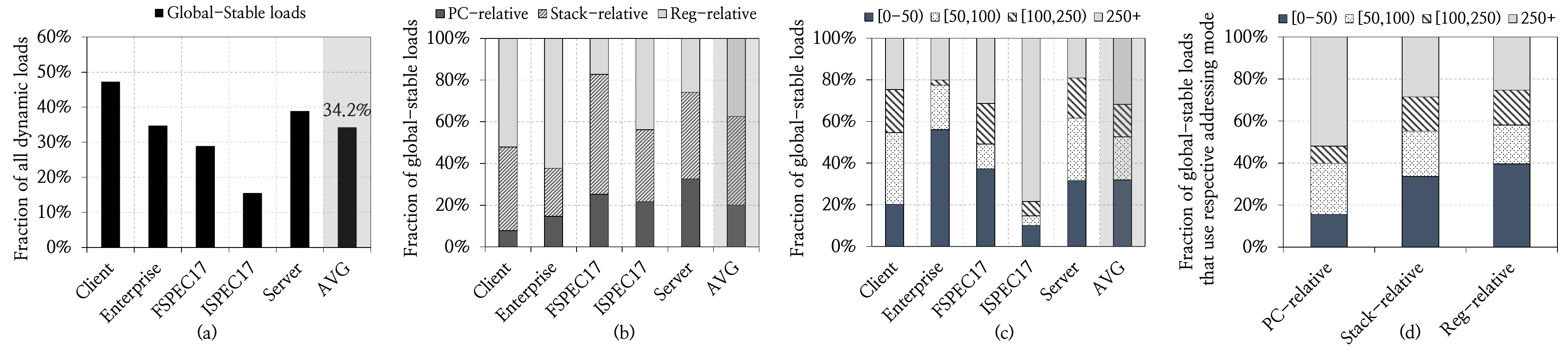}
\caption{
(a) Fraction of dynamic loads that are \rbe{global-stable}. Distribution of \rbe{global-stable} loads by their (b) addressing mode and (b) inter-occurrence distance. (d) Distribution of inter-occurrence distance of \rbe{global-stable} loads from each addressing mode.
}
\label{fig:stable_load_characterization}
\end{figure*}

\subsection{\rbe{Global-Stable} Loads in Real Workloads} \label{sec:headroom_stable_loads_why}

Intuitively, \rbe{global-stable} load instructions would be hard to find in real workloads since such instructions should already be optimized by the compiler. However, we observe that a significant fraction of load instructions in real workloads are \rbe{global-stable} \rbd{even after aggressive compiler optimizations applied}. Fig.~\ref{fig:stable_load_characterization} shows the fraction of \rbc{dynamic} load instructions that are \rbe{global-stable} on average across $90$ workloads divided into five categories. \cref{sec:methodology} discusses our evaluation methodology.
We make two key observations. First, $34.2\%$ of all dynamic loads are \rbe{global-stable}. Second, the fraction of \rbe{global-stable} loads are much higher in \texttt{Client}, \texttt{Enterprise}, and \texttt{Server} workloads as compared to SPEC CPU 2017 workloads (i.e., \texttt{ISPEC17} and \texttt{FSPEC17} categories). \rbd{We} conclude that \rbe{global-stable} load instructions are relatively abundant in real workloads.


\subsubsection{Characterization of \rbe{Global-Stable} Loads} \label{sec:headroom_stable_load_charac}
To understand the source of the \rbe{global-stable} loads in workloads (e.g., accessing variables in global scope, memory accesses in a tight loop),
we further characterize these loads by their addressing mode and the inter-occurrence distance (i.e., the number of instructions between two successive dynamic instances of the same \rbe{global-stable} load instruction).
Fig.~\mbox{\ref{fig:stable_load_characterization}}(b) shows the breakdown of \rbe{global-stable} loads based on their addressing mode. The key takeaway is that \rbe{global-stable} loads use various different addressing modes. On average, $20\%$, $42.6\%$, and $37.4\%$ of all \rbe{global-stable} loads use PC-relative (e.g., loads that access variables in the global scope), stack-relative (i.e., loads that access stack segment using RSP or RBP as their only source register), and register-relative (i.e., loads that use \rbd{other} general-purpose architectural registers as their source) addressing. 
Fig.~\mbox{\ref{fig:stable_load_characterization}}(c) shows the breakdown of \rbe{global-stable} loads based on their inter-occurrence distance. The key takeaway is that \rbe{global-stable} loads have a bimodal inter-occurrence distance distribution. $31.9\%$ of \rbe{global-stable} loads reoccur within $50$ instructions (e.g., loads in a tight loop) on average, whereas $31.8\%$ loads reoccur more than $250$ instructions away (e.g., accessing a global-scope variable across function calls). 
Fig.~\ref{fig:stable_load_characterization}(d) further shows the distribution of inter-occurrence distance of \rbe{global-stable} loads from each addressing mode. 
As we can see, \rbe{global-stable} loads that use PC-relative addressing have long inter-occurrence distance ($52\%$ of these loads have inter-occurrence distance of $250$ or more instructions), whereas \rbe{global-stable} loads that use register-relative addressing have short inter-occurrence distance ($39.6\%$ of these loads have inter-occurrence distance of less than $50$ instructions).

\rbd{We} conclude with three key takeaways. 
First, \rbe{global-stable} load instructions pose diverse characteristics, both in addressing mode and inter-occurrence distance.
Second, the inter-occurrence distance of \rbe{global-stable} loads changes significantly depending on their addressing mode.
Third, an effective load elimination technique should capture elimination opportunities across both short and long inter-occurrence distances.


\setcounter{figure}{4}
\begin{figure*}[!h]
\centering
\includegraphics[scale=0.4]{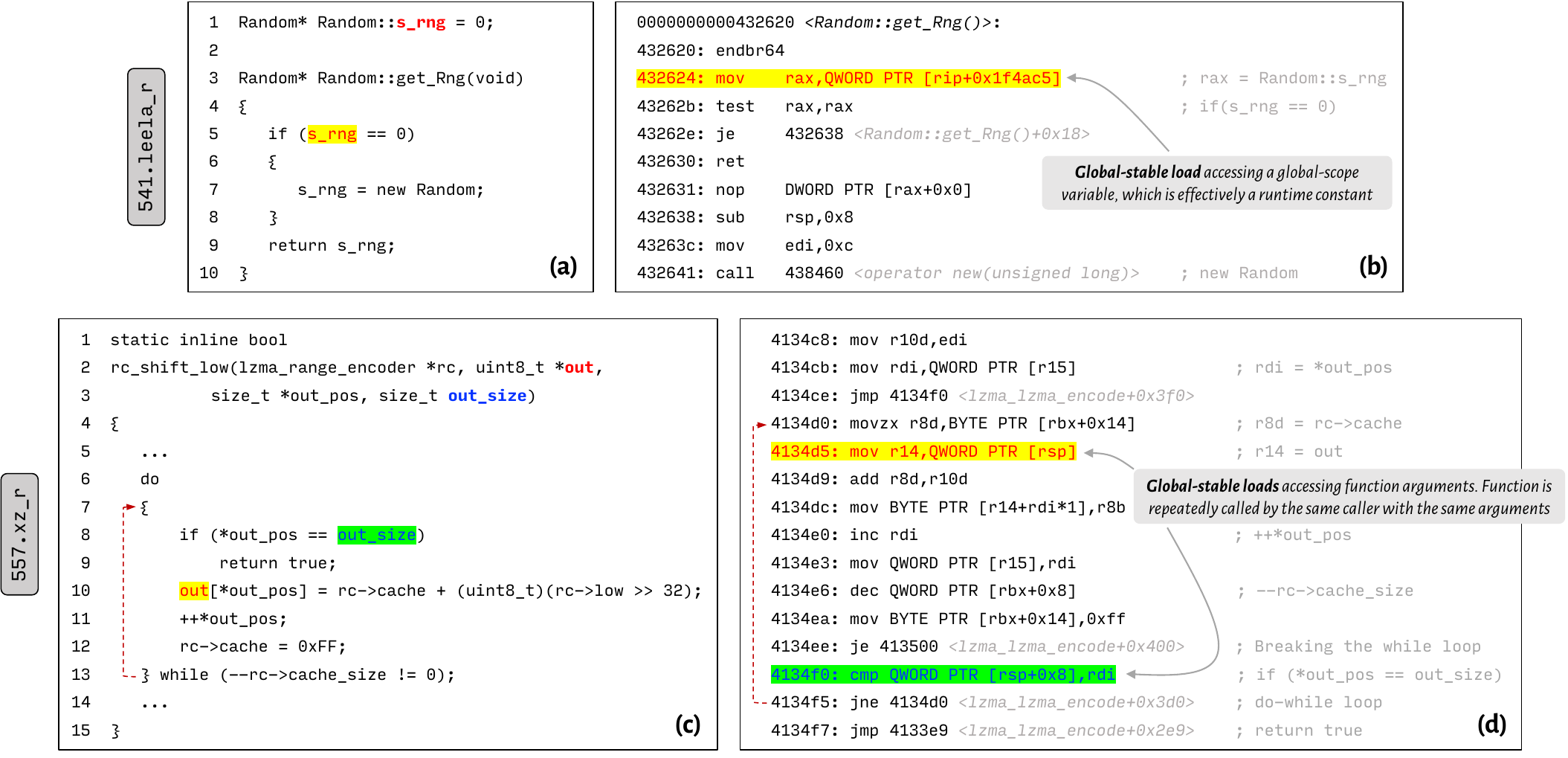}
\caption{
Code \rbd{example} and disassembly from \texttt{541.leela\_r} and \texttt{557.xz\_r} of SPEC CPU 2017 suite. 
}
\label{fig:stable_load_example}
\end{figure*}

\subsection{Why Do \rbe{Global-Stable} Loads Exist?} \label{sec:headroom_workload_analysis}
To understand why \rbd{a} compiler \rbd{with aggressive optimization} \rbd{fails to avoid} \rbe{global-stable} load instructions, we use a custom-made binary instrumentation tool\footnote{We call this tool \emph{Load Inspector}, which is freely available at \url{https://github.com/CMU-SAFARI/Load-Inspector}.} to analyze the disassembly of workload binaries compiled with full optimization \rbd{using} a state-of-the-art off-the-shelf compiler.
Fig.~\ref{fig:stable_load_example}(a) and (b) show a code \rbd{example} from \texttt{541.leela\_r} from SPEC CPU 2017~\cite{spec2017} benchmark suite and its disassembly, respectively. The workload is compiled using the latest GNU g++-13.2 compiler~\cite{gcc_132} at full optimization (i.e., using -O3 flag~\cite{gcc_o3}) for x86-64 instruction set architecture to produce the most optimized binary.  
The highlighted load instruction in Fig.~\ref{fig:stable_load_example} fetches the object pointer \texttt{s\_rng} from memory.
Since \texttt{s\_rng} gets initialized only once at the beginning of the workload, the pointer variable effectively acts as a runtime constant and thus the highlighted load instruction is \rbe{global-stable}. 
The compiler could not eliminate this load instruction since it cannot reserve an architectural register across the global scope of the program to be reused for accessing the \texttt{s\_rng} pointer.



Fig.~\ref{fig:stable_load_example}(c) and (d) show two more examples of \rbe{global-stable} load instructions \rbd{in} a code \rbd{example} from \texttt{557.xz\_r} of SPEC CPU 2017 suite and its disassembly, compiled \rbd{in} the same way as the previous workload. 
\rbd{Each} highlighted load instruction accesses an argument variable to the function \texttt{rc\_shift\_low} that does not change during the function invocation.
Since the function is repeatedly called using the same \rbd{arguments} from the same caller function throughout the workload trace, both the load instructions act as \rbe{global-stable} loads. 
However, as the function \texttt{rc\_shift\_low} gets \emph{inlined} within the body of its caller function (not shown here), the compiler could not allocate architectural registers to store and reuse these variables due to register pressure~\cite{reg_spill}.



We \rbd{conclude} that a state-of-the-art off-the-shelf compiler often fails to optimize \rbe{global-stable} loads due to various empirically-observed reasons, such as accessing runtime constants and local variables of inline functions, \rbd{combined with a limited number of architectural registers}.\footnote{\rbd{We also observe that merely increasing the number of x86-64 architectural registers from $16$ to $32$ has negligible impact on eliminating the global-stable loads at compile time (see ~\cref{sec:arxiv_incr_arch_reg} in the extended version~\cite{constable_extended})}.}

\subsection{Resource Dependence on \rbe{Global-Stable} Loads} \label{sec:headroom_resoruce_hazard}

To quantify the loss of ILP due to resource dependence stemming from \rbe{global-stable} loads, we analyze the utilization of load ports. 
\rbd{Other} hardware resources that are used during a load execution (e.g., RS entry and AGU port) may also cause resource dependence, but we omit them here due to brevity.
Fig.~\mbox{\ref{fig:load_port_util}}(a) shows the fraction of total execution cycles where at least one load port is utilized (we call such cycles \emph{load-utilized}), in our baseline processor\footnote{Our baseline processor has an issue width of six instructions per cycle with three AGU and \rbd{three} load ports, which support a maximum throughput of three loads per cycle.} augmented with a state-of-the-art load value predictor EVES~\cite{eves}.
As we can see, on average $32.7\%$ of the total execution cycles are load-utilized. 
Fig.~\mbox{\ref{fig:load_port_util}}(b) further categorizes the load-utilized cycles of each workload category based on whether or not a \rbe{global-stable} load utilizes a load port.
As we can see, for $23.0\%$ of all load-utilized cycles, a \rbe{global-stable} load takes a load port for its execution, while a non-\rbe{global-stable} load (i.e., a static load instruction that does not fetch the same value from the same load addresses across all dynamic instances) is waiting to be scheduled on the same port.
Had the execution of \rbe{global-stable} loads be eliminated, the non-\rbe{global-stable} loads could have been scheduled faster, which in turn would provide performance benefit.
Thus, we conclude that \rbe{global-stable} load instructions causes significant resource dependence, which can be mitigated by eliminating their execution altogether.

\setcounter{figure}{5}
\begin{figure}[!h] 
\centering
\includegraphics[width=3.4in]{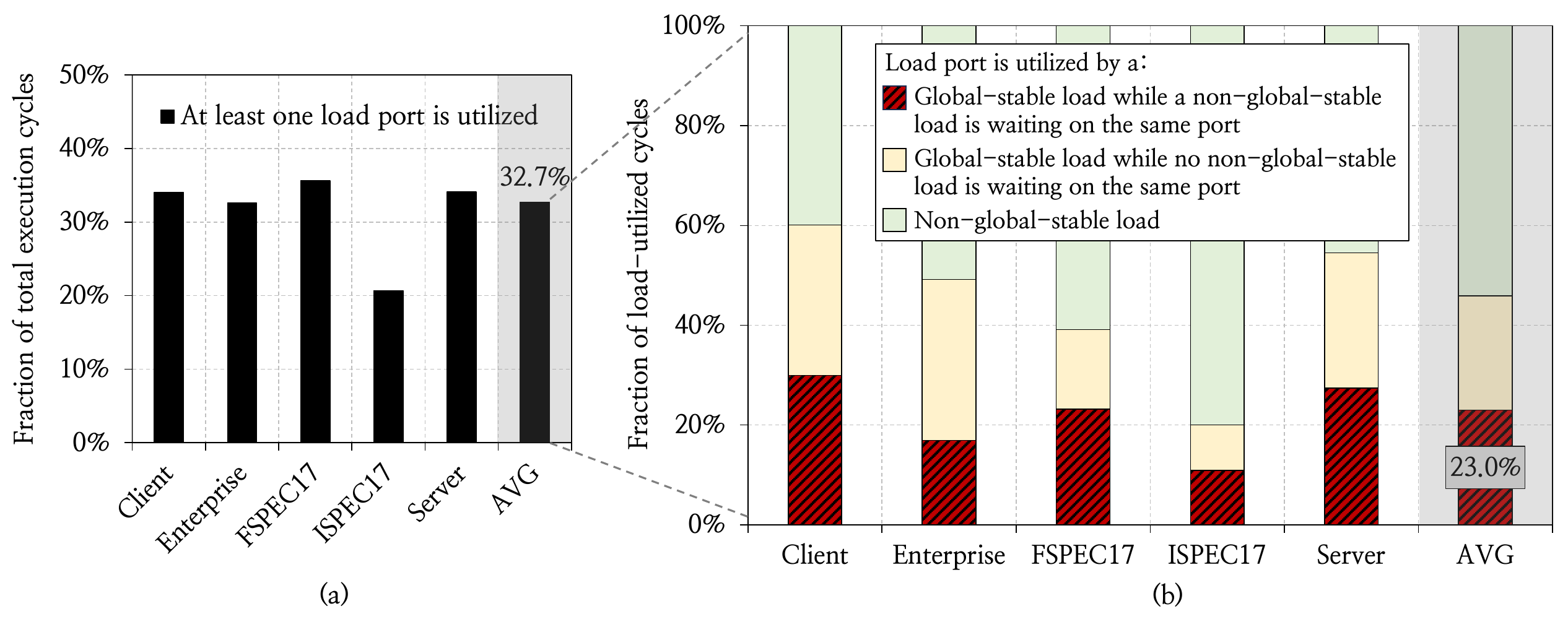}
\caption{(a) Fraction of total execution cycles where at least one load port is utilized (we call such cycles \emph{load-utilized}). (b) Categorization of load-utilized cycles based on whether or not a \rbe{global-stable} load utilizes a load port.}
\vspace{-0.5em}
\label{fig:load_port_util}
\end{figure}

\subsection{Performance Headroom}
\label{sec:headroom_perf_headroom}

\begin{figure}[!h]
\centering
\includegraphics[width=3.4in]{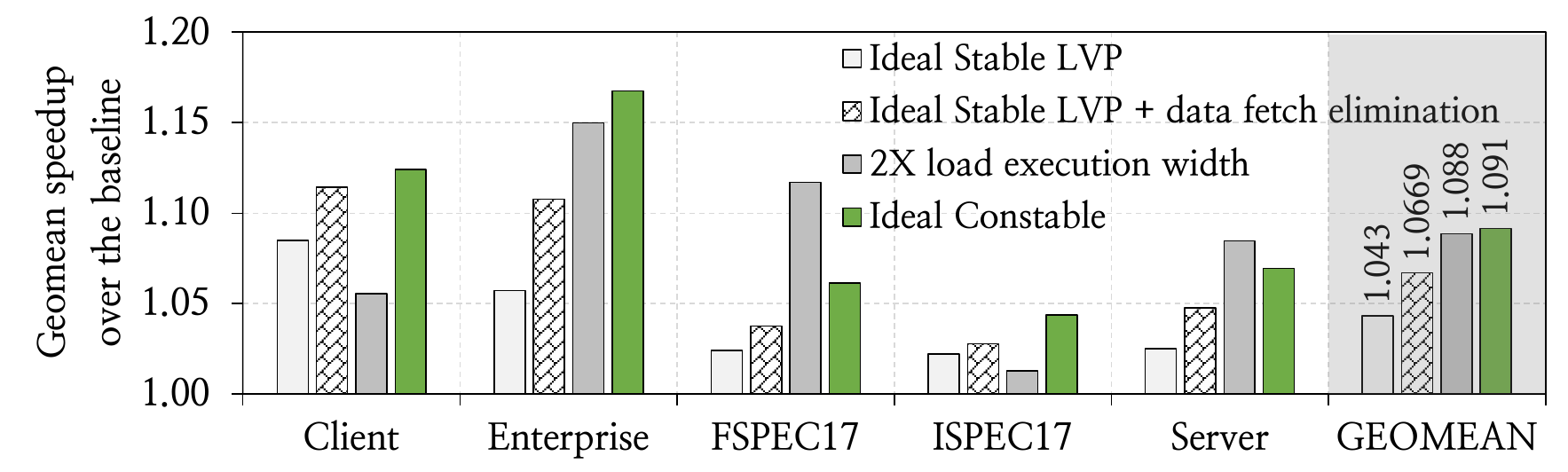}
\caption{Speedup of Ideal Constable against \rbd{Ideal Stable LVP} and a processor with $2\times$ load execution width of the baseline.}
\label{fig:perf_headroom}
\end{figure}

To measure the performance headroom of eliminating \rbe{global-stable} loads, we model an \emph{Ideal Constable} configuration that identifies \rbd{\emph{all}} \rbe{global-stable} load instructions offline and eliminates both component operations of \rbd{their} execution (i.e., load address \rbc{computation} and data \rbc{fetch}).
We also compare Ideal Constable's performance against three other configurations: 
(1) \emph{Ideal \rbd{Stable} LVP}, where \emph{\rbd{all} \rbe{global-stable}} load instructions identified offline are perfectly value predicted, \rbd{and} they are are also executed to verify the predictions, 
(2) \emph{Ideal \rbd{Stable} LVP with data fetch elimination}, where all \rbe{global-stable} load instructions are perfectly value predicted, \rbd{and} the value-predicted loads are executed \rbd{only until the end of} address generation, and 
(3) \emph{$2\times$ load execution width} configuration, where the number of load execution units are doubled \rbd{over} the baseline.
Fig.~\ref{fig:perf_headroom} shows the speedup of each configuration over baseline.

We make four key observations from Fig.~\ref{fig:perf_headroom}. 
First, Ideal Constable provides $9.1\%$ performance improvement on average over the baseline. This shows that eliminating the execution of \rbe{global-stable} loads has high performance headroom. 
Second, Ideal Constable significantly outperforms \rbd{Ideal Stable LVP} ($4.3\%$ on average). This shows that mitigating both data and resource dependence (as done by Ideal Constable) has higher performance potential than only mitigating data dependence (as done by \rbd{Ideal Stable LVP}).
Third, \rbd{Ideal Stable LVP} with data fetch elimination outperforms \rbd{Ideal Stable LVP} ($6.7\%$ on average), yet it falls short to the Ideal Constable. This shows that eliminating \emph{both} the address computation and data \rbc{fetch} operations of a load execution has higher performance potential than just eliminating the data \rbc{fetch}.
Fourth, Ideal Constable even \rbd{slightly} outperforms $2\times$ load execution width configuration, which incurs significantly higher area and power overhead.
\rbd{We} conclude that Constable has significant potential performance benefit by mitigating \emph{both} load data and resource dependence.

\section{Constable: Key Insight} \label{sec:key_idea}

\noindent Constable is based on the \textbf{key insight} that a dynamic instance $I_2$ of a static load instruction $I$ is bound to fetch the same value from the same memory location as the previous dynamic instance $I_1$ of the same static load instruction if the following two conditions are satisfied:
\begin{itemize}
    \item \textbf{Condition~1}: None of the source registers of $I$ has been written between the occurrences of $I_1$ and $I_2$.
    \item \textbf{Condition~2}: No store or snoop request has arrived to the memory address of $I_1$ between the occurrences of $I_1$ and $I_2$. 
\end{itemize}

Satisfying Condition~1 ensures that $I_2$ would have the same load address as $I_1$, and thus the address computation operation of $I_2$ can be safely eliminated.
Satisfying Condition~2 ensures that $I_2$ would fetch the same value from the memory as $I_1$, and thus the data \rbc{fetch operation} of $I_2$ can be safely eliminated.

Constable exploits this observation to operate in two key steps.
First, Constable dynamically identifies load instructions that have repeatedly fetched the same value from the same load address. We call such loads \emph{likely-stable}.
Second, when \rbc{Constable gains enough confidence that a given load instruction is \rbd{likely}-stable}, 
Constable tracks modifications to \rbc{the} source architectural registers \rbc{of the load instruction} and its memory location via two small \rbd{hardware} structures.
Constable \rbc{eliminates} the execution of \rbc{all future instances} of the likely-stable load and \rbd{breaks} the load data dependence using the last-fetched value - until there is a write to \rbd{the} source registers or a store or snoop request to \rbd{the} load address.

\section{Constable: Microarchitecture Design} \label{sec:design}

\subsection{Design Overview}



\begin{figure}[!h] 
\centering
\includegraphics[width=3.4in]{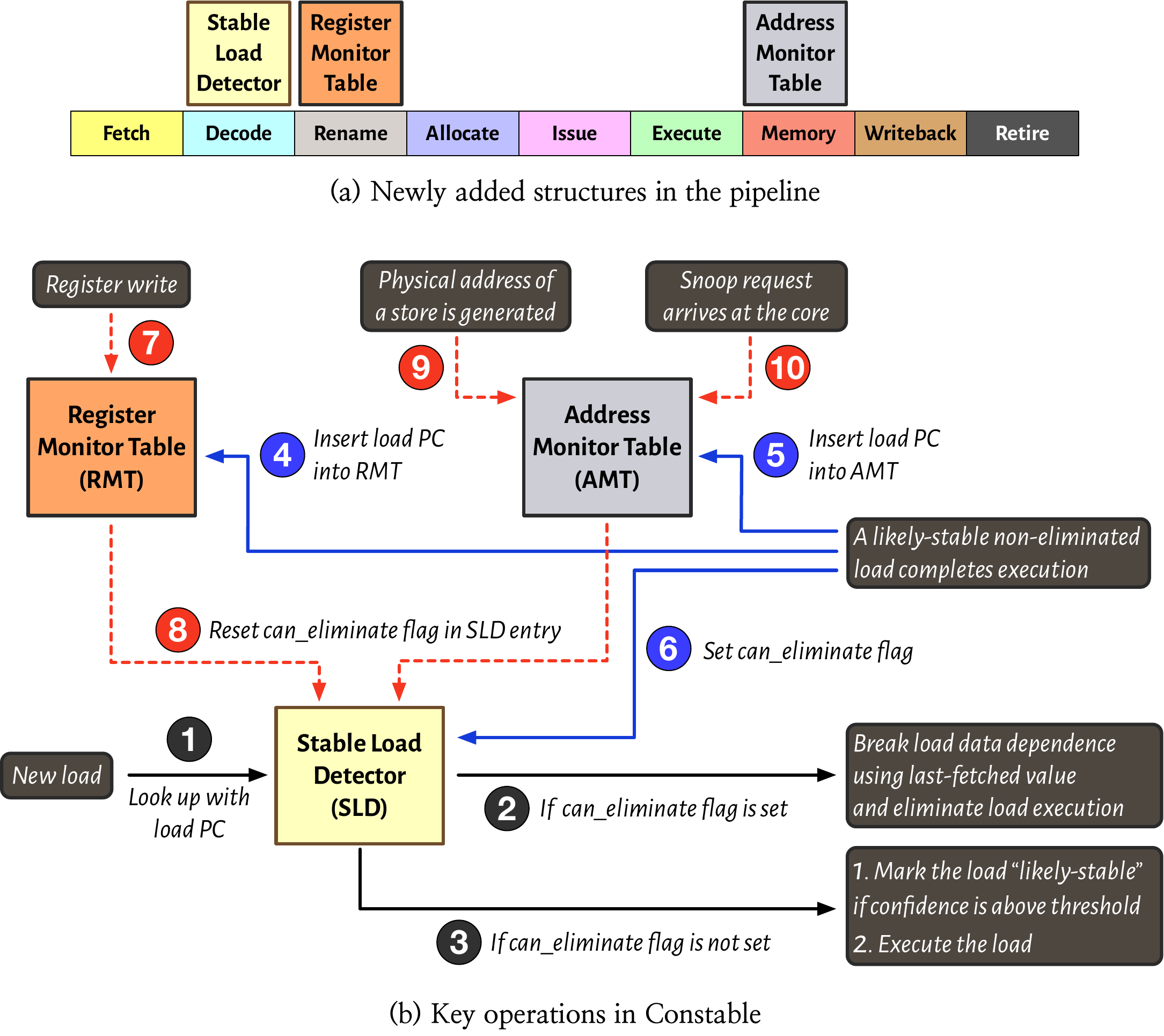}
\caption{Overview of Constable.}
\label{fig:constable_overview}
\end{figure}

Fig.~\ref{fig:constable_overview} shows a high-level overview of Constable. Constable is comprised of three \rbd{main} hardware structures:

\vspace{1pt}
\noindent \textbf{Stable Load Detector (SLD).}
SLD is a program counter (PC)-indexed table that serves three key purposes. 
First, SLD identifies whether or not a given load instruction is likely-stable by \rbd{analyzing} its past dynamic instances. 
Second, SLD \rbd{decides} whether or not the execution of a load instruction can be eliminated.
Third, SLD provides the last-computed load address and the last-fetched data of a given \rbd{likely-stable} load instruction.


\vspace{2pt}
\noindent \textbf{Register Monitor Table (RMT).} 
RMT is an architectural-register-indexed table whose key purpose is to monitor modifications to architectural registers and \rbd{avoid} eliminating a load instruction when its source architectural register gets modified. 
Each RMT entry stores a list of load PCs that are currently getting eliminated that use the corresponding architectural register as their source. 
In the rename stage, every instruction looks up RMT using its destination architectural register and resets the elimination status of any load PC from the corresponding RMT entry in SLD to ensure that any future instances of that load instruction will not be eliminated. In essence, RMT \rbd{enforces} the Condition~1 for eliminating a load instruction (\cref{sec:key_idea}).

\vspace{2pt}
\noindent \textbf{Address Monitor Table (AMT).}
AMT is a physical-address-indexed table whose key purpose is to monitor modifications in the memory and \rbd{avoid} eliminating a load instruction when the memory location from which it fetches the data gets modified.
Each AMT entry stores a list of load PCs that are currently getting eliminated that access the corresponding physical memory address. 
Every store or snoop request looks up AMT using its physical address and resets the elimination status of any load PC from the corresponding AMT entry in SLD to ensure that any subsequent instances of that load will not be eliminated further. In essence, AMT \rbd{enforces} the Condition~2 for eliminating a load instruction (\cref{sec:key_idea}).

\subsection{Identifying Likely-Stable Loads} \label{sec:design_identifying_likey_stable_loads}

SLD employs a confidence-based learning mechanism to identify likely-stable load instructions based on the execution \rbd{outcomes} of their past dynamic instances. Each SLD entry stores four key pieces of information: (1) last-computed load address, (2) last-fetched value, (3) a $5$-bit \emph{stability confidence \rbd{level}} and (4) a \texttt{can\_eliminate} flag that represents whether or not an instance of this load instruction can be eliminated. 
When a non-eliminated load instruction completes execution in the writeback stage, Constable checks the SLD using the load PC to compare the last-computed load address and last-fetched value with the current load address and value. If both the address and value match, Constable increments the stability \rbd{confidence level} by one; otherwise, it halves the confidence. If the stability \rbd{confidence level} surpasses a threshold (set to $30$ in our evaluation), Constable identifies subsequent load instances from the same PC as likely-stable.


\subsection{Eliminating Load Execution} \label{sec:design_break_dep}

During the rename stage, a load instruction first checks the SLD using the load PC (\circled{1} in Fig.~\ref{fig:constable_overview}). 
If the \texttt{can\_eliminate} flag is set in the corresponding SLD entry, Constable breaks the load data dependence using the last-fetched value stored in the SLD entry and eliminates its execution (\circled{2}). 
If the \texttt{can\_eliminate} flag is not set, Constable checks the stability \rbd{confidence level} stored in the SLD entry. If the \rbd{confidence level} is above threshold, Constable marks the load instruction as likely-stable and executes it normally as the baseline (\circled{3}). 
Only a load instruction marked as likely-stable can set the \texttt{can\_eliminate} flag during the writeback stage of its execution (see \cref{sec:design_update_load_complete}).


\vspace{2pt}
\noindent \textbf{Microarchitecture for breaking load data dependence.}
Breaking load data dependence requires supplying the load value to all dependent in-flight instructions. Prior works on LVP achieve this by writing the value to the physical register file (PRF) or to a separate value table~\cite{rami_ap}. Since writing to PRF either requires adding expensive write ports to PRF~\cite{perais2014eole,perais2015bebop,orosa2018avpp} or a latency-sensitive arbitration of the existing write ports~\cite{perais2021ssr,rami_ap}, Constable implements load data dependence breaking using a small extra register file (only $32$ entries), called xPRF, which is dedicated to hold the values of the in-flight eliminated load instructions.\footnote{
We implement Constable using xPRF as having a small PRF to break data dependence has been shown to be more area- and energy-efficient than adding new write ports to the existing PRF~\cite{rami_ap}. However, Constable can also be implemented by adding or arbitrating PRF write ports. For a fair evaluation, we also implement the LVP and MRN techniques considered in this work using xPRF. As such, xPRF is not considered as an additional structure for Constable.
}

If SLD decides to eliminate the load execution, Constable stores the last-fetched value provided by SLD in an available xPRF register and converts the load instruction into a three-operand register move instruction: the source \rbd{is} the xPRF register, the destination \rbd{is} the destination architectural register of the load, and the third operand \rbd{is} the last-computed load address provided by the SLD. If there is no available xPRF register, Constable does not eliminate the load and executes it normally as the baseline. 
We observe this happens rarely \rbd{(only in $0.2\%$ of the instances)} in our evaluation with a $32$-entry xPRF. 
In the rename stage, the converted register move instruction simply maps its destination register to the source xPRF register to complete its execution (similar to move elimination~\cite{trace_cache,cont_opt}). 
\rbd{Doing so} enables the dependents of the converted register move instruction to get scheduled by reading the xPRF register value. 
In the allocation stage, the converted register move instruction allocates a reorder buffer (ROB) entry and a load buffer (LB) entry. 
The address field in the LB entry gets updated with the last-computed load address embedded within the move instruction as the third operand. 
This address field in the LB entry is later required to correctly disambiguate the eliminated load from the in-flight stores~\cite{mem_disambig} as discussed in \cref{sec:design_in_flight_stores}. 
Since the execution of the converted register move instruction has already been completed in the rename stage, the instruction bypasses the remaining pipeline stages \rbd{and resources} directly \rbd{to retirement} based on the in-order retirement logic.

\subsection{Updating Constable Structures}

\subsubsection{\textbf{Updates When a Likely-Stable Non-Eliminated Load Finishes Execution}} \label{sec:design_update_load_complete}
During the writeback stage of the pipeline, when a likely-stable yet not eliminated load finishes its execution, Constable updates its structures to eliminate subsequent instances of the same load instruction. This happens in three steps. 
First, Constable looks up RMT with its source architectural registers. For each source register, Constable inserts the load PC into the corresponding RMT entry (\bluecircled{4}).
Second, Constable looks up AMT with the physical address of the load instruction. If the load address is found, Constable inserts the load PC into the corresponding AMT entry (\bluecircled{5}).
If the address is not found, Constable inserts a new AMT entry for the load address and inserts the load PC into the new AMT entry. 
Third, Constable looks up SLD with the load PC and sets the \texttt{can\_eliminate} flag of the corresponding entry (\bluecircled{6}). Setting the \texttt{can\_eliminate} flag allows Constable to eliminate the execution of subsequent instances of the same load instruction.

\subsubsection{\textbf{Updates during Register Renaming}}
In the rename stage, Constable checks the destination architectural register of every instruction and updates its structures to \rbd{avoid} eliminating subsequent instances of any load instruction that uses the destination register as its source.
This happens in two steps.
First, Constable looks up RMT with the architectural destination register of every instruction (\redcircled{7}).
If there is any load PC in the corresponding RMT entry, Constable looks up the SLD using each load PC and resets the \texttt{can\_eliminate} flag in the corresponding entry in SLD (\redcircled{8}).

\subsubsection{\textbf{Updates on a Store Instruction}} \label{sec:design_update_store}
When the address of a store instruction gets generated, Constable updates its structures to \rbd{avoid} eliminating subsequent instances of any load instruction that fetches data from the same memory address as the store. 
This happens in two steps.
First, Constable looks up AMT using the physical store address (\redcircled{9}). 
If the address is found in AMT, Constable looks up SLD using each load PC in the AMT entry and resets the \texttt{can\_eliminate} flag from the corresponding entry in SLD (\redcircled{8}).
Second, after resetting \texttt{can\_eliminate} flag for all load PCs in the AMT entry, Constable evicts the AMT entry.


\subsubsection{\textbf{Updates on a Snoop Request}} \label{sec:design_update_snoop}
To safely eliminate loads in multi-core systems, Constable monitors snoop requests coming to the core and updates its structures to \rbd{avoid} eliminating subsequent instances of any load instruction that fetches data from the same memory address as the snoop.
Constable handles \rbd{a} snoop \rbd{request} in a similar way as a store request. 
When a snoop request arrives at the core, Constable looks up AMT using the snoop address (\redcircled{10}). 
If the address is found, Constable looks up SLD using each load PC in the AMT entry and resets the \texttt{can\_eliminate} flag from the corresponding entry in SLD (\redcircled{8}).
Finally, Constable evicts the AMT entry.

\subsection{Disambiguating Eliminated Loads \\ from In-Flight Stores} \label{sec:design_in_flight_stores}
When a store instruction computes its address, Constable accesses AMT and resets the \texttt{can\_eliminate} flag for all load instructions accessing the same memory location (see \cref{sec:design_update_store}). 
This prevents Constable from eliminating any \emph{subsequent occurrences} of those load instructions. 
However, in a processor that aggressively issues loads out-of-order~\cite{mem_disambig,franklin1996arb,moshovos1997dynamic}, there may be eliminated loads in the pipeline that are younger than the store instruction and whose addresses match with the store address. 
We observe that this happens \rbd{rarely (see~\cref{sec:arxiv_ext_eval_pipe_flush} in the extended version~\cite{constable_extended})} since Constable considers a load instruction to be eligible for elimination only if it meets the stability \rbd{confidence level} threshold.
In such infrequent cases, Constable exploits the existing memory disambiguation logic~\cite{mem_disambig,data_preload} that matches the store address with the address of every load in the LB. If a violation is caught, Constable flushes the pipeline and \rbd{re-executes all younger instructions, including the incorrectly-eliminated load} (see the example in \cref{sec:design_working_example}).




\subsection{Maintaining Coherence in Multi-Core Systems} \label{sec:design_multi_core_coherence}

Constable relies on monitoring snoop requests for tracking modifications in the memory by other processor cores to safely eliminate loads in a multi-core system. However, monitoring snoop requests poses the following two key challenges.

\vspace{1pt}
\noindent \textbf{Loss of elimination opportunity due to clean eviction.}
In a multi-core system with a directory-based coherence protocol~\cite{censier1978new}, when a cacheline gets evicted from a core-private cache, the core-valid bit (CV-bit) corresponding that core \rbd{(i.e., the \emph{own core})} gets reset in the directory entry of that cacheline~\cite{scalable_directory,directory_evaluation,scd,gupta1992reducing}. Since resetting CV-bit prevents the directory from sending any further snoop request to that cacheline to the core, on every core-private cache eviction, Constable needs to \rbd{avoid} eliminating any load instruction that accesses the evicted cacheline. This poses two key drawbacks. First, if the evicted cacheline is clean (e.g., eviction due to limited cache capacity or cache conflict), Constable loses elimination opportunity \rbd{(we quantify the impact of such elimination opportunity loss in~\cref{sec:arxiv_eval_clean_eviction} in~\cite{constable_extended})}. Second, for every core-private cache eviction, Constable needs to look up and invalidate the corresponding AMT entry, which increases design complexity. 
To address these drawbacks, we propose to \emph{pin} the own core's CV-bit of a cacheline that is accessed by an \rbd{eliminated} load instruction.
When the memory request of a likely-stable yet not eliminated load returns from the cache hierarchy,
Constable pins the own core's CV-bit in the directory entry of that cacheline.
Pinning CV-bit ensures that (1) the coherence protocol would send any snoop request to that cacheline to the own core, even if the cacheline gets clean-evicted from the core-private cache, and (2) Constable does not need to look up AMT on every core-private cache eviction.
The CV-bit is reset as soon as a snoop request is delivered to the core, as per the normal directory-based coherence protocol.

\vspace{1pt}
\noindent \textbf{Tracking snoop requests \rbd{at} cacheline address granularity.}
Unlike a store instruction that contains a full memory address, a snoop request contains a cacheline address. 
Thus, to support AMT lookup using a snoop address, Constable indexes AMT using physical addresses 
at cacheline granularity. 
\rbd{This} may \rbd{cause loss of} elimination opportunities due to false address collisions (e.g., a store to a cacheline may reset the \texttt{can\_eliminate} flag of a load instruction that accesses different bytes of the same cacheline accessed by the store).
However, we find that the performance impact of such elimination opportunity loss is negligible. 
Constable with a cacheline-address-indexed AMT \rbd{has only} $0.4\%$ \rbd{lower average} performance than a \rbd{Constable with} full-address-indexed AMT.
This is primarily because the compiler tends to lay out memory addresses accessed by likely-stable load instructions together (e.g., a group of function arguments laid out in the same cacheline of stack memory segment), which reduces \rbd{the overhead of false address collisions}.


\subsection{Other Design Decisions} \label{sec:design_design_decisions}

\subsubsection{\textbf{Architecting SLD}} \label{sec:design_design_decisions_sld}
Designing SLD with sufficient read/write ports is crucial for realizing Constable's performance \rbd{benefits}.
Constable reads SLD for every load instruction to identify likely-stable loads in the rename stage (\mbox{\circled{1}} in Fig.~\mbox{\ref{fig:constable_overview}}). Thus SLD needs to support the read bandwidth of the expected number of load instructions in a group of instructions getting renamed together in every cycle (we call this a \emph{rename group}).\footnote{We model a six-wide rename architecture (see \cref{sec:methodology_perf}).} We observe that a rename group contains $1.93$ loads on average across all workloads, and $98.3\%$ of all rename groups have less than or equal to two loads. 
\rbd{Thus,} we model SLD with three read ports. If there are more than three loads in a rename group, we stall the rename stage until Constable finishes SLD lookup for every load in that group.

\begin{figure}[!h] 
\centering
\includegraphics[width=3.4in]{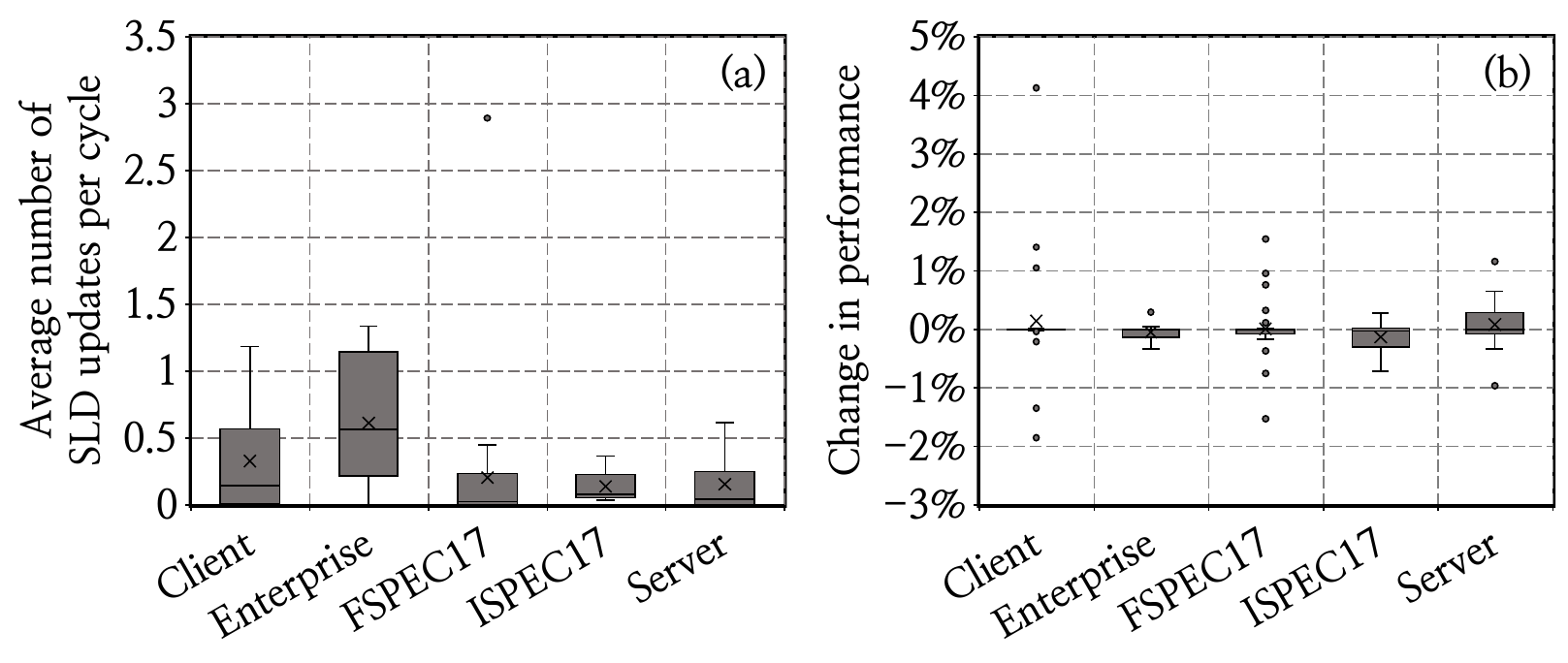}
\caption{(a) \rbd{Average} number of SLD updates per cycle during rename stage. (b) \rbd{Change} in performance when Constable's structures are updated only by correct path instructions vs. all instructions without \rbd{updating} \rbd{Constable's} structures on branch misprediction recovery.}
\label{fig:design_decisions}
\end{figure}

Constable may need to update the \texttt{can\_eliminate} flag in SLD on every RMT update, which happens for each instruction in a rename group (\redcircled{7} and \redcircled{8}). Since each RMT entry may contain a list of likely-stable load PCs, the expected number of SLD updates per cycle can vary in a large range. 
Fig.~\mbox{\ref{fig:design_decisions}}(a) shows the average number of observed SLD updates per cycle for every workload as a box-and-whiskers plot.\footnote{
Each box is lower- (upper-) bounded by the first (third) quartile. The box size represents the inter-quartile range (IQR). The whiskers extend to 1.5$\times$IQR range on each side, and the cross-marked values in the box show the mean.
}
As we can see, we observe only $0.28$ SLD updates per cycle on average across all workloads. $98.23\%$ of all cycles on average across all workloads have two or \rbd{fewer} SLD updates. This is because, at any point in time, only a small fraction of all load PCs \rbd{($14.7\%$ on average)} satisfies the stability \rbd{confidence level} threshold \rbd{in order} to be tracked by RMT entries. Thus, we model SLD with two write ports. If there are more than two SLD updates in a cycle, we stall the rename stage until Constable finishes SLD update for every load instruction in that rename group.

\subsubsection{\textbf{Handling Wrong Path Execution}} \label{sec:design_design_decisions_wrong_path}
In presence of branch prediction, Constable's structures may get updated by wrong path instructions (especially, steps~\redcircled{7} and~\redcircled{8} in Fig.~\mbox{\ref{fig:constable_overview}}). This may result in an unnecessary loss of elimination opportunity, unless the structures are restored on a branch misprediction recovery. 
To understand the need for restoring Constable's structures, we measure the change in performance of Constable when its structures are updated only by the instructions on the correct path against when they are updated by all instructions without \rbd{an update} mechanism on branch misprediction recovery, and show it as a box-and-whiskers plot in Fig.~\mbox{\ref{fig:design_decisions}}(b). 
The key observation is that $82$ out of $90$ workloads show less than $1\%$ absolute change in performance, while the average performance change is only $0.2\%$. Thus, we model Constable without \rbd{any update} mechanism for its structures on a branch misprediction recovery.

\begin{figure*}[!h]
\centering
\includegraphics[scale=0.27]{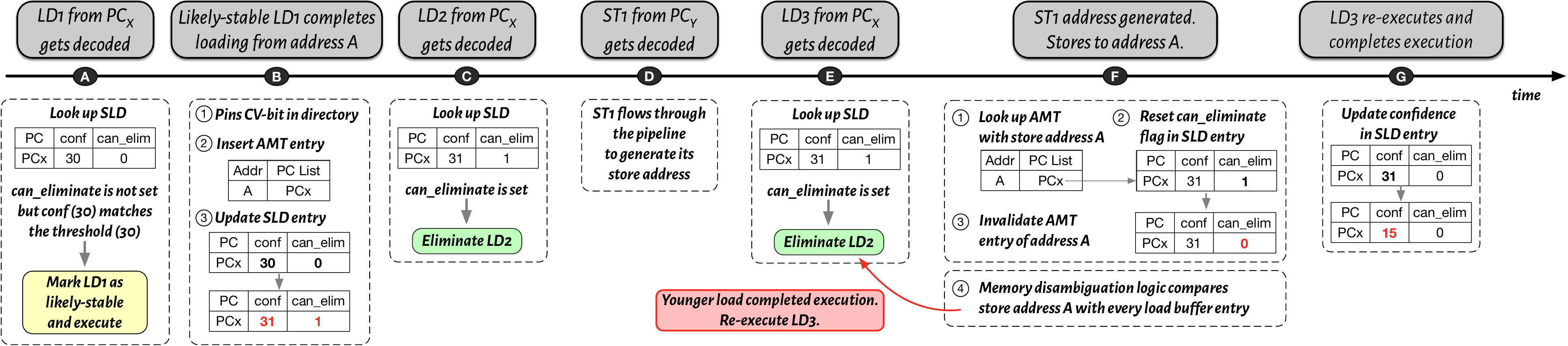}
\caption{An illustrative example of \rbd{Constable's operation}.}
\label{fig:working_example}
\end{figure*}







\subsubsection{\textbf{Handling Changes in Physical Address Mapping}}
AMT monitors memory locations accessed by all \rbd{eliminated} load instructions in physical address space. 
This poses a challenge: when the physical memory mapping changes, the physical memory address tracked by an AMT entry may not be associated with the corresponding eliminated load anymore. In that case, to \rbd{avoid} incorrectly eliminating load execution, 
Constable resets the \texttt{can\_eliminate} flag of all SLD entries and invalidates all RMT and AMT entries when the physical memory mapping changes (e.g., context switch).



\subsection{An Illustrative Example} \label{sec:design_working_example}
To put it together, Fig.~\ref{fig:working_example} illustrates an example of \rbd{Constable's operation}. For this example, we consider that the loads $LD_1$, $LD_2$, and $LD_3$ are three dynamic instances of the static load instruction $LD$ with a PC value $PC_x$ and the source registers of $LD$ do not get modified between $LD_1$ and $LD_3$.
We also assume the stability \rbd{confidence level} threshold is set to $30$. 

When $LD_1$ gets decoded (\circled{A} in Fig.~\ref{fig:working_example}), Constable checks SLD and finds that the stability \rbd{confidence level} of $PC_x$ matches the threshold, yet the \texttt{can\_eliminate} flag is not set.
In this case, Constable marks $LD_1$ as likely-stable and executes it normally. 
When $LD_1$ finishes its execution (\circled{B}), Constable (1) pins the CV-bit corresponding to the own core in the coherence directory entry of address A, (2) updates AMT and RMT (not shown here), and (3) increments the stability \rbd{confidence level} in SLD. Since $LD_1$ is marked as likely-stable, Constable sets the \texttt{can\_eliminate} flag in SLD entry.
When $LD_2$ gets decoded (\circled{C}), Constable eliminates executing $LD_2$ since the \texttt{can\_eliminate} flag is set.
Now a store $ST_1$ from a different $PC_y$ gets decoded (\circled{D}). This store instruction would ultimately modify the memory address touched by $LD$. 
However, before $ST_1$ could generate its store address, $LD_3$, which is younger than $ST_1$ in program order, gets decoded (\circled{E}) and Constable \emph{incorrectly} eliminates its execution since the \texttt{can\_eliminate} flag is set. 
When $ST_1$ finally generates its store address (\circled{F}), Constable resets the \texttt{can\_eliminate} flag to prevent eliminating subsequent instances of $LD$ and evicts the AMT entry. 
However, the existing memory disambiguation logic probes the load buffer with the store address and finds out that a younger load $LD_3$ has been incorrectly completed. 
As a result, the memory disambiguation logic aborts and re-executes $LD_3$ \rbd{(and all instructions younger than $LD_3$ that are not shown here)}. \rbd{When $LD_3$ completes its execution} (\circled{G}), \rbd{it} halves the stability \rbd{confidence level} counter. 

\subsection{Storage Overhead} \label{sec:design_storage_overhead}
Table~\ref{table:storage_overhead} shows the storage overhead of Constable. 
Constable requires only $12.4$ KB storage \rbd{per core of the processor (see~\cref{sec:methodology_perf})}.

\begin{table}[htbp]
  \centering
  \scriptsize
  \caption{Storage overhead of Constable. \\ \footnotesize{(Our baseline system models a 48-bit physical address space)}}
    \begin{tabular}{L{4.5em}||m{22em}||L{3.5em}}
    \thickhline
    \Tstrut \textbf{Structure} \Bstrut & \Tstrut \textbf{Description} \Bstrut & \Tstrut \textbf{Size} \Bstrut\\
    \thickhline
    \Tstrut \textbf{SLD} &
    \Tstrut \begin{minipage}{22em}
      \scriptsize
      \vskip 2pt
      \begin{itemize}[leftmargin=1em]
        \setlength\itemsep{-0.01em}
        \item \# entries: $512$ ($32$ sets $\times$ $16$ ways)
        \item Entry size: tag ($24b$) + addr ($32b$) + val ($64b$) + \rbd{confidence level} ($5b$) + can\_eliminate flag ($1b$)
      \end{itemize}
      \vskip 2pt
    \end{minipage} \Bstrut &
    \Tstrut \textbf{7.9~KB} \Bstrut \\
    \hline
    \Tstrut \textbf{RMT} \Bstrut &
    \Tstrut \begin{minipage}{22em}
      \scriptsize
      \vskip 2pt
      \begin{itemize}[leftmargin=1em]
        \setlength\itemsep{-0.01em}
        \item $16$ load PCs for each stack registers (RSP and RBP)
        \item $8$ load PCs for each remaining $14$ architectural registers in x86-64
      \end{itemize}
      \vskip 2pt
    \end{minipage} \Bstrut &
    \Tstrut \textbf{0.4~KB} \Bstrut \\
    \hline
    \Tstrut \textbf{AMT} \Bstrut &
    \Tstrut \begin{minipage}{22em}
      \scriptsize
      \vskip 2pt
      \begin{itemize}[leftmargin=1em]
        \setlength\itemsep{-0.01em}
        \item \# entries: 256 ($32$ sets $\times$ $8$ ways)
        \item Entry size: physical address tag ($32b$) + \# hashed load PCs ($4\times24b$)
      \end{itemize}
      \vskip 2pt
    \end{minipage} \Bstrut &
    \Tstrut \textbf{4.0~KB} \Bstrut \\
    \thickhline
    \Tstrut \textbf{Total} \Bstrut & & \Tstrut \textbf{12.4~KB}\Bstrut \\
    \thickhline
    \end{tabular}%
  \label{table:storage_overhead}%
\end{table}%
\section{Differences from Related Prior Works} \label{sec:key_prior_works}

The key idea of early-executing the address computation and/or eliminating the data \rbc{fetch} operations of a load instruction has been explored in many prior works. 
We divide such works into the following four \rbd{categories} to better understand Constable's differences from them.

\vspace{1pt}
\noindent \textbf{Prior works that early execute address computation.} 
Prior works \rbd{propose} speculative and non-speculative techniques to early or fast execute address computation of a load instruction. 
\cite{austin_load_addr} uses a fast, speculative carry-free addition to speed up load address computation.
\cite{austin1995zero} caches the values of recently-used registers to speculatively compute load address.
Early load address resolution (ELAR)~\cite{elar} tracks stack register \rbd{values} using a small computation unit in the decode stage to \emph{safely and non-speculatively} compute the load address of most stack loads immediately after \rbd{the} decode stage. 
Register file prefetching (RFP)~\cite{rfp} predicts the load address of an instruction to prefetch its data to the register file.

Constable differs from these works in one major way. 
These prior works necessitate \rbd{the execution of} the load instruction whose load address has been computed early. 
Works that \emph{speculatively} compute load address~\cite{austin_load_addr,austin1995zero,rfp} need to execute the load to verify the speculation. ELAR, which employs a safe technique to \emph{non-speculatively} compute address of stack loads, still needs to fetch the load data from the memory hierarchy. 
Constable safely eliminates \emph{both the address computation and the data \rbc{fetch} operations} of a load execution altogether.
We evaluate Constable against ELAR and  RFP in~\cref{sec:eval_prior_work} demonstrating its performance \rbd{benefits}.

\vspace{1pt}
\noindent \textbf{Prior work that eliminates \rbd{the} data \rbc{fetch} operation.}
Lipasti et al. (we \rbd{call} it \rbd{the} Lipasti load value predictor~\cite{lipasti1996value} or LLVP) observe that some static load instructions have highly-predictable load values, which they \rbd{call} \emph{constant loads}. 
LLVP proposes a microarchitecture to bypass the data \rbc{fetch} \rbd{operations} of constant loads.
LLVP maintains addresses of constant loads in a table called constant verification unit (CVU) and invalidates a CVU entry by \rbd{observing} a store request to the corresponding address.
To verify the predicted value of a constant load, LLVP first computes its load address. If the address is found in CVU, LLVP bypasses the data \rbc{fetch} operation.

Constable differs from LLVP in one major way. 
LLVP advocates for eliminating \emph{\rbd{only}} the data \rbc{fetch} operation of a value-predicted constant load. 
Constable, on the other hand, eliminates both the address computation and data \rbc{fetch} operations of a load instruction execution. 
As~\cref{sec:headroom_perf_headroom} shows, eliminating both address computation and data \rbc{fetch} operations (as done by Ideal Constable) \rbd{provides} higher performance benefit than eliminating only data \rbc{fetch} (as done by Ideal Stable LVP with data fetch elimination).

\vspace{1pt}
\noindent \textbf{Prior works that enable memoization.}
Memoization~\cite{michie1968memo}, caches computed results from \rbd{repeated} code executions, \rbd{enabling a} program or a microarchitecture to skip redundant computations when encountering identical input sets.
Memoization has been applied in both software~\cite{michie1968memo,richardson_function_memo,conners1999compiler,suresh2015intercepting,suresh2017compile} and in hardware at various program granularities, including instruction-level~\cite{harbison1982architectural,richardson1993exploiting,citron1998accelerating,sodani1997dynamic,sodani1998empirical,molina1999dynamic}, basic-block-level~\cite{huang1999exploiting}, trace-level~\cite{gonzalez1999trace}, and function-level~\cite{citron2000hardware}.
Early works on instruction-level memoization aim to accelerate long-latency operations (e.g., floating point multiplication and division) by storing their operands and results in value caches~\cite{richardson1993exploiting,oberman1996reducing}.
Sodani and Sohi propose a PC-indexed \emph{reuse buffer} to store the results of (multiple) dynamic instances of every static instruction~\cite{sodani1997dynamic}.
Molina et al. improve upon the reuse buffer to capture reuse of results across dynamic instances of different static instructions~\cite{molina1999dynamic}.

Constable, \rbd{in principle}, resembles instruction-level memoization with three key differences that make Constable more performant, lightweight, and usable in today's high-performance multi-core processors.
First, prior works aim to memoize (multiple) results of \emph{every} static instruction, irrespective of whether or not the results would be useful for instruction elimination. This requires a large memoization buffer, often as large as L1 data cache~\cite{citron2002revisiting,zhang2017leveraging,zhang2017leveraging2}, to capture elimination opportunities across long inter-occurrence distances (see \cref{sec:headroom_stable_load_charac}). Gonzalez et al. have shown that, while such a large memoization buffer may provide a significant performance benefit, the benefits reduce significantly when the latency to access the buffer is considered~\cite{gonzalez1998performance}.
Constable on the other hand (a) only targets loads, and (b) employs a confidence-based mechanism to filter out likely-stable load instructions from all loads.
This significantly reduces Constable's storage overhead and design complexity (e.g., port requirements of SLD as discussed in \cref{sec:design_design_decisions_sld}) while providing high elimination coverage.
\rbd{Second, prior works may delay retrieving the memoized instruction output until its source register values are available~\cite{harbison1982architectural,richardson1993exploiting,citron1998accelerating,molina1999dynamic}.
As a result, these works may not alleviate resource dependence on hardware structures like the reservation station.}
\rbd{Constable, however, explicitly monitors changes in source architectural registers of likely-stable load instructions and eliminates them early in the pipeline, alleviating resource dependence from both load reservation station and load execution unit.}
Third, prior works may not be applied in today's high-performance multi-core processors as they do not address challenges related to (a) keeping memoization buffer coherent across multiple cores, and (b) maintaining program correctness in presence of out-of-order load issue. Constable addresses both these challenges (\cref{sec:design_in_flight_stores} and \cref{sec:design_multi_core_coherence}) and we extensively verify its correctness via functional simulation (\cref{sec:methodology_func_veri}).



\vspace{1pt}
\noindent \textbf{Prior works \rbd{on dynamic instruction} optimization.} 
Prior works on trace-cache-based optimizations~\cite{trace_cache,trace_processor} and rePLay framework~\cite{replay,replay2} enable a wide range of runtime code optimizations (e.g., move elimination, zero-idiom elimination) that can eliminate instructions in microarchitecture. 
Continuous optimization (CO)~\cite{cont_opt} builds \rbd{on} these works and enables removing redundant instructions (including loads) using the register renaming logic.

Constable differs from these works in two key ways.
First, these schemes learn optimizations offline per-trace (or frame) basis. Constable learns optimization online and applies directly to the program's dynamic instruction stream.
Second, \rbd{unlike Constable}, CO \rbd{does not} eliminate a load instruction in a multi-core system \rbd{in order to maintain the coherency}.
Our baseline system already implements many runtime optimizations to non-memory instructions at renaming stage  (see~\cref{sec:methodology}).
\section{Methodology} \label{sec:methodology}

\subsection{Performance Modeling} \label{sec:methodology_perf}
We evaluate Constable using \rbd{an} in-house, cycle-accurate, industry-grade simulator that simultaneously runs both functional and microarchitectural simulation on a workload. 
We faithfully model a $6$-wide out-of-order processor core configured similar to the Intel Golden Cove~\cite{goldencove,goldencove_microarch,goldencove_microarch2,spr1} as our baseline. 
Table~\ref{table:sim_params} shows the key microarchitectural parameters.
\rbd{We include} MRN and various dynamic optimizations in the rename stage of the baseline processor, as highlighted in bold.
For a comprehensive analysis, we evaluate Constable and other competing mechanisms on the baseline system, both without SMT (called \emph{noSMT}) and with $2$-way SMT (called \emph{SMT2}). 
For noSMT \rbd{configuration}, all hardware resources inside core are fully available to the single running software context.
For SMT2 \rbd{configuration}, resources inside core (including Constable) are either statically-partitioned or dynamically-shared between both software contexts~\cite{amd_smt}.
Unless stated otherwise, all reported results are from noSMT simulations.

\begin{table}[htbp]
  \centering
  \scriptsize
  \caption{Simulation parameters.\\\footnotesize{(IDQ: Instruction Decode Queue, SB: Store Buffer)}}
    \begin{tabular}{p{4em}||p{28em}}
    \thickhline
    \Tstrut \textbf{Basic} \Bstrut & \Tstrut x86-64 core clocked at 3.2~GHz with 2-way SMT support \Bstrut \\
    \hline
    \vskip 1pt \textbf{Fetch \& Decode} \Bstrut & \Tstrut 8-wide fetch, TAGE/ITTAGE branch predictors~\cite{seznec2006case}, 20-cycle misprediction penalty, 32KB 8-way L1-I cache, 4K-entry 8-way micro-op cache, 6-wide decode, 144-entry IDQ, loop-stream detector~\cite{lsd} \Bstrut \\
    \hline
    \vskip 4pt \textbf{Rename} \Bstrut & \Tstrut 6-wide, 288 integer, 220 512-bit and 320 256-bit physical registers, \textbf{Memory Renaming~\cite{mrn}, zero elimination~\cite{trace_cache}, move elimination~\cite{trace_cache,cont_opt}, constant folding~\cite{trace_cache,cont_opt}, branch folding~\cite{branch_folding}} \Bstrut \\
    \hline
    \textbf{Allocate} & \Tstrut 512-entry ROB, 240-entry LB, 112-entry SB, 248-entry RS \\
    \hline
    \vskip 4pt \textbf{Issue \& Retire} \Bstrut & \Tstrut 6-wide issue to 12 execution ports; 5, 3, 2, and 2 ports for ALU, load, store-address, and store-data execution. Port 0, 1, and 5 are used for vector instructions. Aggressive out-of-order load scheduling with memory \rbd{dependence} prediction~\cite{store_set,moshovos1997streamlining}, 6-wide retire\Bstrut \\
    \hline
    \vskip 10pt \textbf{Caches} \Bstrut & \Tstrut \textbf{L1-D}: 48KB, 12-way, 5-cycle latency, LRU, PC-based stride prefetcher~\cite{stride}; \textbf{L2}: 2MB, 16-way, 12-cycle round-trip latency, LRU, stride + streamer~\cite{streamer} + SPP~\cite{spp}; \textbf{LLC}: 3MB, 12-way, 50-cycle data round trip latency~\cite{goldencove_microarch,hermes}, dead-block-aware replacement policy~\cite{sdbp}, streamer, MESIF~\cite{mesif} protocol\Bstrut \\
    \hline
    \vskip 0.2pt \textbf{Memory} \Bstrut & \Tstrut 4 channels, 2 ranks/channel, 8 banks/rank, 2KB row-buffer/rank, 64b bus/channel, DDR4, tCAS=22ns, tRCD=22ns, tRP=22ns, tRAS=56ns\Bstrut \\
    \hline
    \hline
    \Tstrut \textbf{Optional} \Bstrut & 
    \Tstrut \begin{minipage}{28em}
      \scriptsize
      \vskip 2pt
      \begin{itemize}[leftmargin=0.8em]
        \setlength\itemsep{-0.01em}
        \item \textbf{EVES}~\cite{eves}: exact design of the CVP-1 \textbf{32KB} storage track winner
        \item \textbf{ELAR}~\cite{elar}: with additional adder and RSP register in decode stage
        \item \textbf{RFP}~\cite{rfp}: $2$K-entry prefetch table, $64$-entry page address table, $128$-entry RFP-inflight table. Total size: \textbf{12.5KB}
        \item \textbf{Constable} (this work). Total size: \textbf{12.4KB}
      \end{itemize}
    \end{minipage} \\
    \thickhline
    \end{tabular}%
  \label{table:sim_params}%
\end{table}%

\subsection{Power Modeling} \label{sec:methodology_power}
We use \rbd{an} in-house, RTL-validated power model to measure the dynamic power consumption \rbd{of} the core. 
We report the overall core power consumption by breaking it down into four key units: (1) front end (FE), (2) out-of-order (OOO), (3) non-memory execution unit (EU), and (4) memory execution unit (MEU). We further break down the OOO power into three sub-units: RS, Register Alias Table (RAT), and ROB. We also break down the MEU power into two sub-units: L1-D cache, and data translation look-aside buffer (DTLB).
To faithfully model the power consumption of Constable, we estimate the read/write access energy and leakage power of Constable's structures using CACTI 7.0~\cite{cacti} $22$nm library. We scale the estimates to $14$nm technology using~\cite{tech_scaling} to make the estimates compatible with our core power model. Table~\ref{table:area_power} shows the access energy, leakage power, and the area estimate of Constable's structures. We report RMT and SLD power in the RAT component and AMT power in the L1-D component of the core power model.


\begin{table}[htbp]
  \centering
  \scriptsize
  \caption{Access energy, leakage power, and area \rbd{estimates} of Constable's structures in 14nm technology.}
    \begin{tabular}{m{6.2em}||C{5.5em}C{5.5em}C{6em}C{2.8em}}
    \thickhline
    \Tstrut \textbf{Component}\Bstrut & \Tstrut \textbf{Read access energy (pJ)}\Bstrut & \Tstrut \textbf{Write access energy (pJ)}\Bstrut & \Tstrut \textbf{Leakage power (mW)}\Bstrut & \Tstrut \textbf{Area (mm2)}\Bstrut\\
    \thickhline
    \Tstrut \textbf{SLD} ($7.9$KB, $3$R/$2$W ports) \Bstrut & \Tstrut$10.76$\Bstrut & \Tstrut$16.70$\Bstrut & \Tstrut$1.02$\Bstrut  & \Tstrut$0.211$\Bstrut \\
    \hline
    \Tstrut \textbf{RMT} ($0.4$KB, $2$R/$6$W ports) \Bstrut & \Tstrut$0.15$\Bstrut  & \Tstrut$0.20$\Bstrut  & \Tstrut$0.31$\Bstrut  & \Tstrut$0.004$\Bstrut \\
    \hline
    \Tstrut \textbf{AMT} ($4.0$KB, $1$R/$1$W ports) \Bstrut & \Tstrut$1.58$\Bstrut  & \Tstrut$4.22$\Bstrut  & \Tstrut$0.74$\Bstrut  & \Tstrut$0.017$\Bstrut \\
    \thickhline
    \end{tabular}%
  \label{table:area_power}%
\end{table}%

\subsection{Workloads} \label{sec:methodology_workloads}

We evaluate Constable using $90$ workload traces that span across a diverse set of $58$ workloads. Our workload suite contains all benchmarks from the SPEC CPU 2017 suite~\cite{spec2017}, and many well-known \texttt{Client}, \texttt{Enterprise}, and \texttt{Server} workloads.
Each trace contains a snapshot of the processor and the memory state (1) to drive both the functional and microarchitecture simulation \rbd{models}, and (2) to faithfully simulate wrong-path execution.
Each trace is carefully selected to be representative of the overall workload.
Table~\ref{table:workloads} summarizes the complete list of workloads.


\begin{table}[htbp]
  \centering
  \scriptsize
  \caption{Workloads used for evaluation.}
    \begin{tabular}{m{3em}C{6em}C{3.5em}m{16em}}
    \thickhline
    \Tstrut \textbf{Suite} & \Tstrut \textbf{\#Workloads} & \Tstrut \textbf{\#Traces} & \Tstrut \textbf{Example Workloads} \\
    \thickhline
    \Tstrut \textbf{Client} & \Tstrut 16    & \Tstrut 22    & \Tstrut DaCapo~\cite{dacapo}, SYSmark~\cite{sysmark}, TabletMark~\cite{tabletmark}, JetStream2~\cite{jetstream}\\
    \hline
    \Tstrut \textbf{Enterprise} & \Tstrut 9    & \Tstrut 14    & \Tstrut SPECjEnterprise~\cite{specjenterprise}, SPECjbb~\cite{specjbb}, LAMMPS~\cite{lammps} \\
    \hline
    \Tstrut \textbf{FSPEC17} & \Tstrut 13     & \Tstrut 29    & \Tstrut All from SPECrate FP 2017~\cite{spec2017} \\
    \hline
    \Tstrut \textbf{ISPEC17}  & \Tstrut 10    & \Tstrut 11    & \Tstrut All from SPECrate Integer 2017~\cite{spec2017} \\
    \hline
    \Tstrut \textbf{Server}    & \Tstrut 10    & \Tstrut 14    & \Tstrut Hadoop~\cite{hadoop}, Linpack~\cite{linpack}, Snort~\cite{snort}, BigBench~\cite{bigbench} \\ 
    \thickhline
    \end{tabular}%
  \label{table:workloads}%
\end{table}%

\subsection{Evaluated Mechanisms}

For a comprehensive analysis, we evaluate Constable standalone and in combination with three prior works: (1) a state-of-the-art load value predictor EVES~\cite{eves}, (2) early load address resolution (ELAR)~\cite{elar}, and (3) register file prefetching (RFP)~\cite{rfp}.
For EVES, we use the optimized implementation that won the first championship value prediction (CVP-1) in \rbd{the} $32$~KB storage budget track~\cite{cvp1}.
For ELAR, we follow the same microarchitecture design as proposed in~\cite{elar}.
For RFP, we sweep and select the configuration parameter values that provide the highest performance benefit over the baseline. EVES, RFP, and Constable apply their optimizations to load instructions with data size up to $64$ bits. Table~\ref{table:sim_params} shows the \rbd{overheads} of all evaluated mechanisms. 



\subsection{Functional Verification of Constable} \label{sec:methodology_func_veri}

Since Constable completely eliminates a load instruction in microarchitectural simulation, we cannot verify the functional correctness of Constable in the same way as LVP or MRN techniques. \rbd{To} functionally verify Constable, we enforce a \emph{golden check} at the \rbd{retirement} stage of every load instruction.
The golden check matches the load address and the load data from the functional simulation model with those from the microarchitectural simulation model. In case of a mismatch, the golden check aborts the simulation. We extensively verify the functional correctness of Constable using a broader set of $3400$ traces and ensure that no single trace fails the simulation.

\section{Evaluation} \label{sec:evaluation}

\subsection{Performance Improvement} \label{sec:eval_perf}

\subsubsection{\textbf{noSMT \rbd{Configuration}}} \label{sec:eval_perf_no_smt}
Fig.~\ref{fig:perf_main} shows the geomean performance of EVES, Constable, and Constable and Ideal Constable (\cref{sec:headroom_perf_headroom}) combined with EVES normalized to the baseline for each workload category. We make three key observations.
First, Constable alone improves performance by $5.1\%$ on average over the baseline, which is similar to the performance improvement of EVES ($4.7\%$ on average) while incurring $\frac{1}{2}\times$ of \rbd{EVES'} storage overhead.
Second, when combined with EVES, Constable improves performance by $8.5\%$ on average over the baseline, which is $3.7\%$ \rbd{higher} than EVES alone.
Third, when combined with EVES, Constable provides $82.9\%$ of the performance improvement provided by Ideal Constable.

\begin{figure}[!h] 
\centering
\includegraphics[width=3.4in]{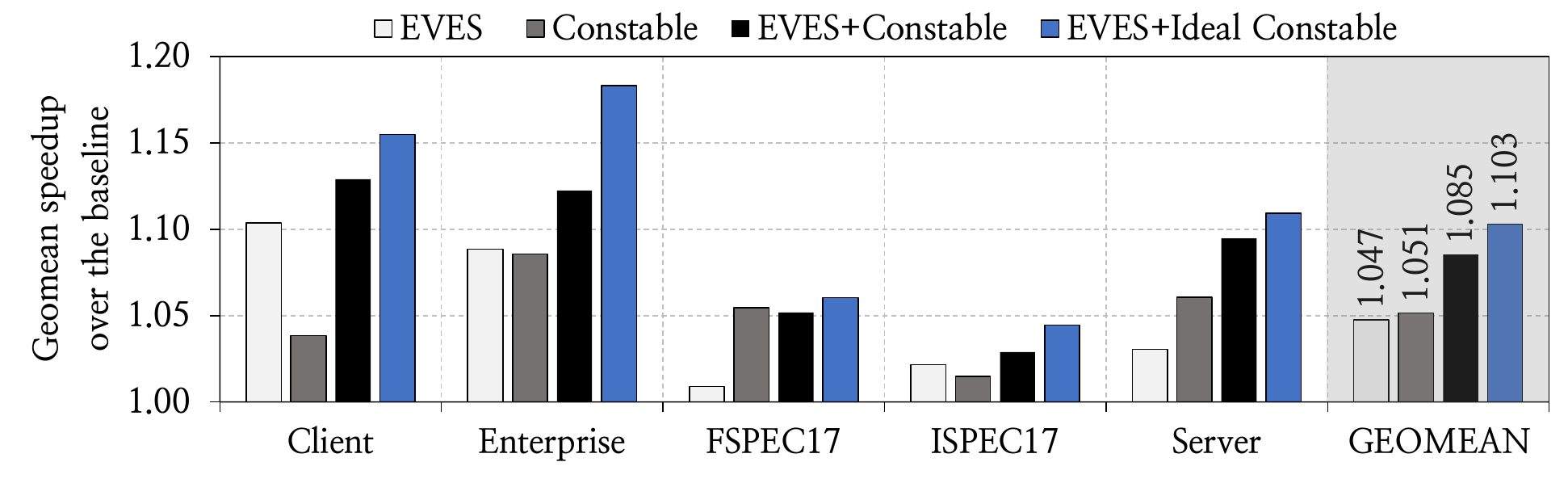}
\caption{Speedup over the baseline \rbd{(noSMT)}.}
\label{fig:perf_main}
\end{figure}

\begin{figure}[!h] 
\centering
\includegraphics[width=3.4in]{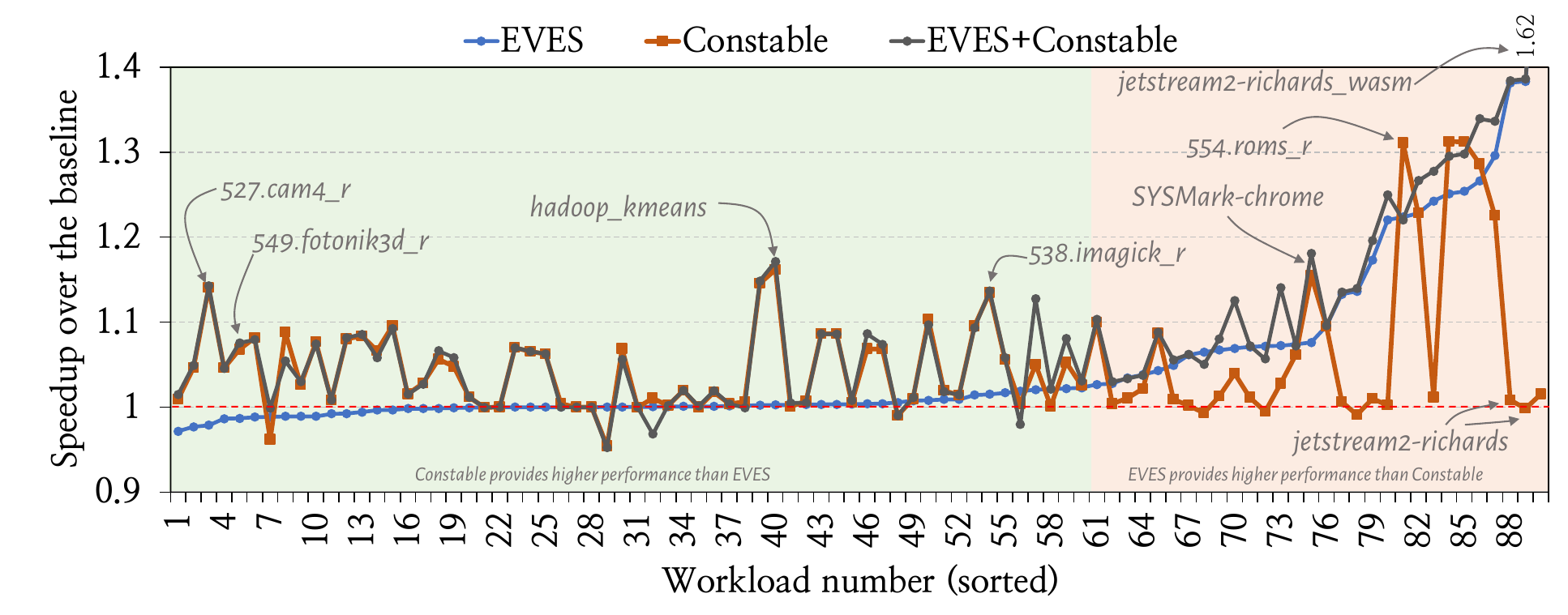}
\caption{\rbd{Speedup} of all workloads \rbd{(noSMT)}.}
\label{fig:perf_line}
\end{figure}

\vspace{1pt}
\noindent \textbf{Per-workload performance.} To better understand Constable's performance improvement, Fig.~\ref{fig:perf_line} shows the performance line graph of EVES, Constable, and Constable combined with EVES for every workload. \rbd{Workloads} are sorted in ascending order of the performance gain of EVES over the baseline. We make three key observations.
First, Constable outperforms EVES by $4.9\%$ on average in $60$ \rbd{of the} $90$ workloads (highlighted in green). In the remaining 30 workloads (highlighted in red), EVES outperforms Constable by $9.2\%$ on average.
Second, Constable combined with EVES consistently outperforms both EVES and Constable alone in every workload.

\begin{figure}[!h] 
\centering
\includegraphics[width=3.4in]{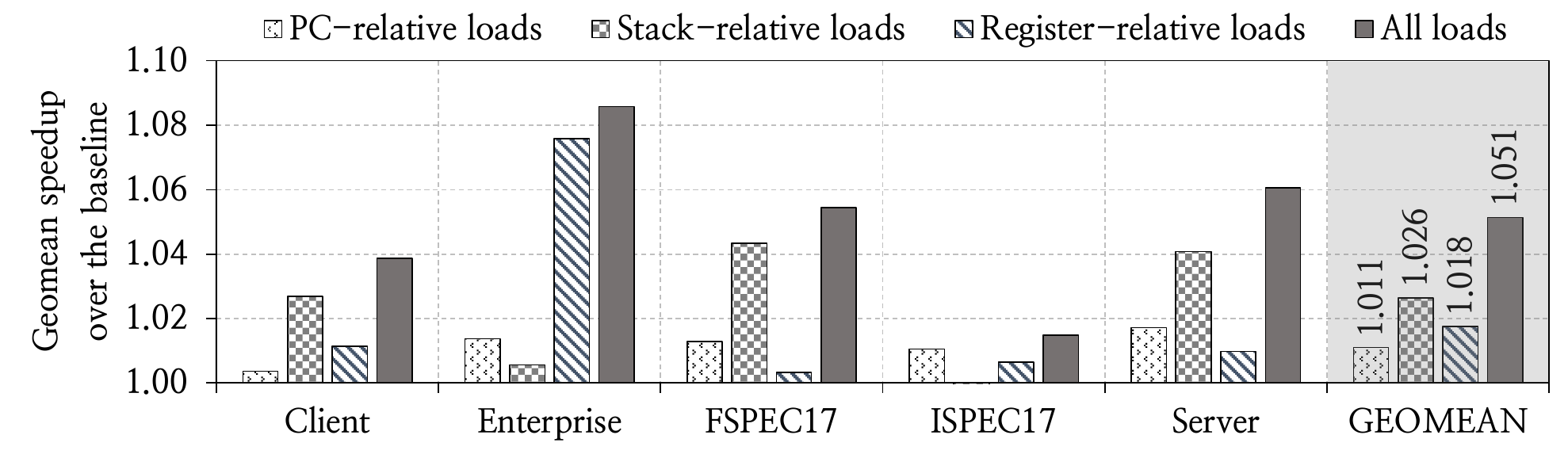}
\caption{Speedup of Constable by eliminating execution of only PC-relative, stack-relative, and register-relative loads.}
\label{fig:perf_breakdown}
\end{figure}

\vspace{1pt}
\noindent \textbf{Load category-wise performance.} 
To understand the performance \rbd{benefits} contributed by different load categories, Fig.~\mbox{\ref{fig:perf_breakdown}} compares the geomean performance of Constable when it eliminates only PC-relative, stack-relative, and register-relative loads with that of a full-blown Constable. 
The key takeaway is that each three individual types of loads contribute towards Constable's overall performance benefit. Eliminating only PC-relative, stack-relative, and register-relative loads provide a performance improvement of $1.1\%$, $2.6\%$, and $1.8\%$, respectively, which nearly get added up to a $5.1\%$ improvement by the full-blown Constable.

Based on these results, we conclude that (1) Constable provides a significant performance benefit over a wide range of workloads both by itself and when combined with a state-of-the-art load value predictor EVES, and (2) Constable's performance benefit comes from eliminating all types of loads.

\subsubsection{\textbf{SMT2 \rbd{Configuration}}} \label{sec:eval_perf_smt}

Fig.~\ref{fig:perf_main} shows the geomean performance of EVES, Constable, and Constable combined with EVES normalized to the baseline. We make two key observations.
First, unlike \rbd{in} noSMT \rbd{configuration}, Constable significantly outperforms EVES in the baseline \rbd{with} SMT2. Constable alone improves performance by $8.8\%$ on average over the baseline, whereas EVES alone improves performance by $3.6\%$. This is because, unlike EVES, Constable's load elimination fundamentally reduces utilization of load execution resources, which face increased contention in presence of SMT.
Second, combining Constable with EVES continues to provide additional performance benefit than EVES alone. Constable with EVES improves performance by $11.3\%$ on average over the baseline in SMT2.
\rbd{We} conclude that Constable provides even more performance benefit in presence of SMT as compared to non-SMT system, \rbd{due to high resource contention in SMT systems}.


\begin{figure}[!h] 
\centering
\includegraphics[width=3.4in]{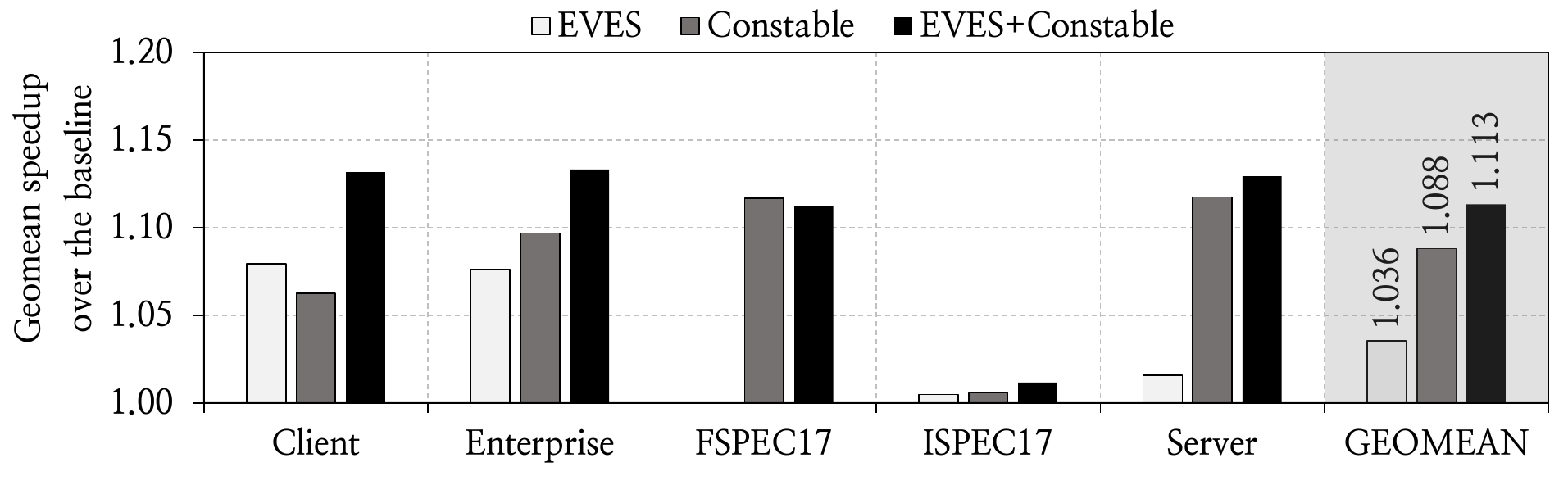}
\caption{Speedup over the baseline (SMT2).}
\label{fig:perf_main}
\end{figure}

\subsection{Performance Comparison with Prior Works} \label{sec:eval_prior_work}

\begin{figure}[!h] 
\centering
\includegraphics[width=3.4in]{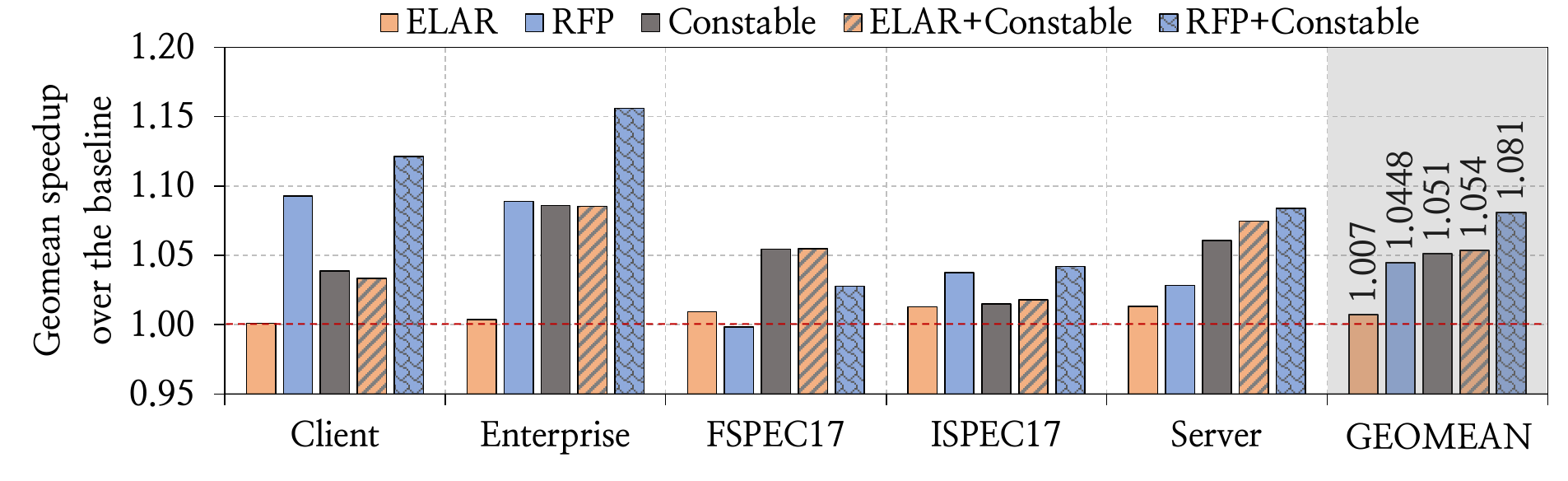}
\caption{Speedup of Constable \rbd{over} ELAR and RFP.}
\label{fig:prior_work}
\end{figure}

Fig.~\ref{fig:prior_work} shows the geomean performance of Constable standalone and when combined with ELAR and RFP in the baseline. We make two key observations.
First, Constable alone outperforms both ELAR and RFP. ELAR, RFP, and Constable \rbd{improve} performance on average by $0.74\%$, $4.4\%$ and $5.1\%$, respectively over the baseline. ELAR provides relatively small performance benefit over baseline as our baseline already implements constant folding~\cite{trace_cache,cont_opt}, which can track stack register modifications in the form of $RSP \leftarrow RSP \pm immediate$ before the execution stage. 
Second, when combined with ELAR and RFP, Constable provides \rbd{more} performance benefit than ELAR and RFP alone, respectively. This shows that Constable can be applied \rbd{along with} these proposals to provide even more performance benefit.

\subsection{Loads \rbd{Eliminated by Constable}} \label{sec:evaluation_load_coverage}

\begin{figure}[!h] 
\centering
\includegraphics[width=3.4in]{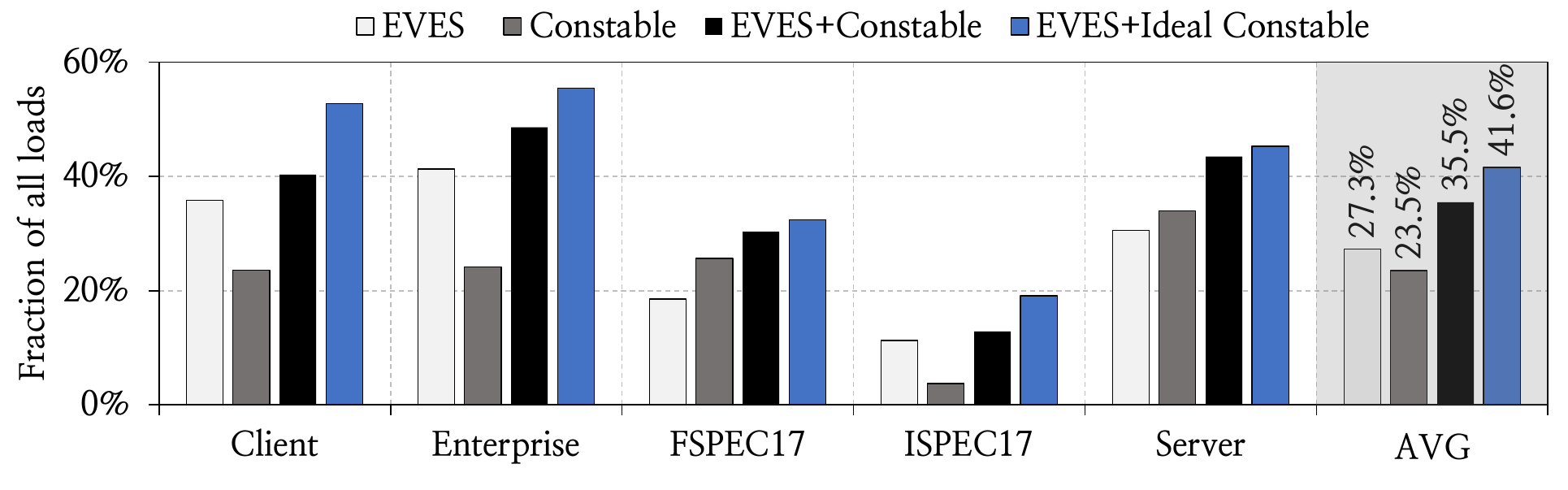}
\caption{Load coverage of Constable \rbd{versus} EVES.}
\label{fig:cov_main}
\end{figure}

Fig.~\ref{fig:cov_main} shows the load coverage (i.e., the fraction of load instructions that are either eliminated or value-predicted by Constable or EVES, respectively) of EVES, Constable, and Constable and Ideal Constable combined with EVES in the baseline system. We make three key observations.
First, Constable alone covers $23.5\%$ of the loads, whereas EVES covers $27.3\%$. 
This is because Constable target loads which show \emph{both} value and address locality (i.e., repeatedly fetching same value from same memory address), whereas EVES target loads that show only value locality. 
Despite \rbd{its} lower coverage, Constable matches performance of EVES as Constable mitigates \emph{both} data dependence and resource dependence on covered loads. 
Second, Constable combined with EVES has higher load coverage ($35.5\%$ on average) than EVES alone.
Third, when combined with EVES, Constable provides $85.4\%$ of the coverage of Ideal Constable.
We conclude that Constable covers a significant fraction of the load instructions both by itself and combined with EVES.

\vspace{2pt}
\subsubsection{Coverage of \rbe{Global-Stable} Loads} \label{sec:evaluation_stable_load_coverage}
\begin{figure}[!h] 
\centering
\includegraphics[width=3.4in]{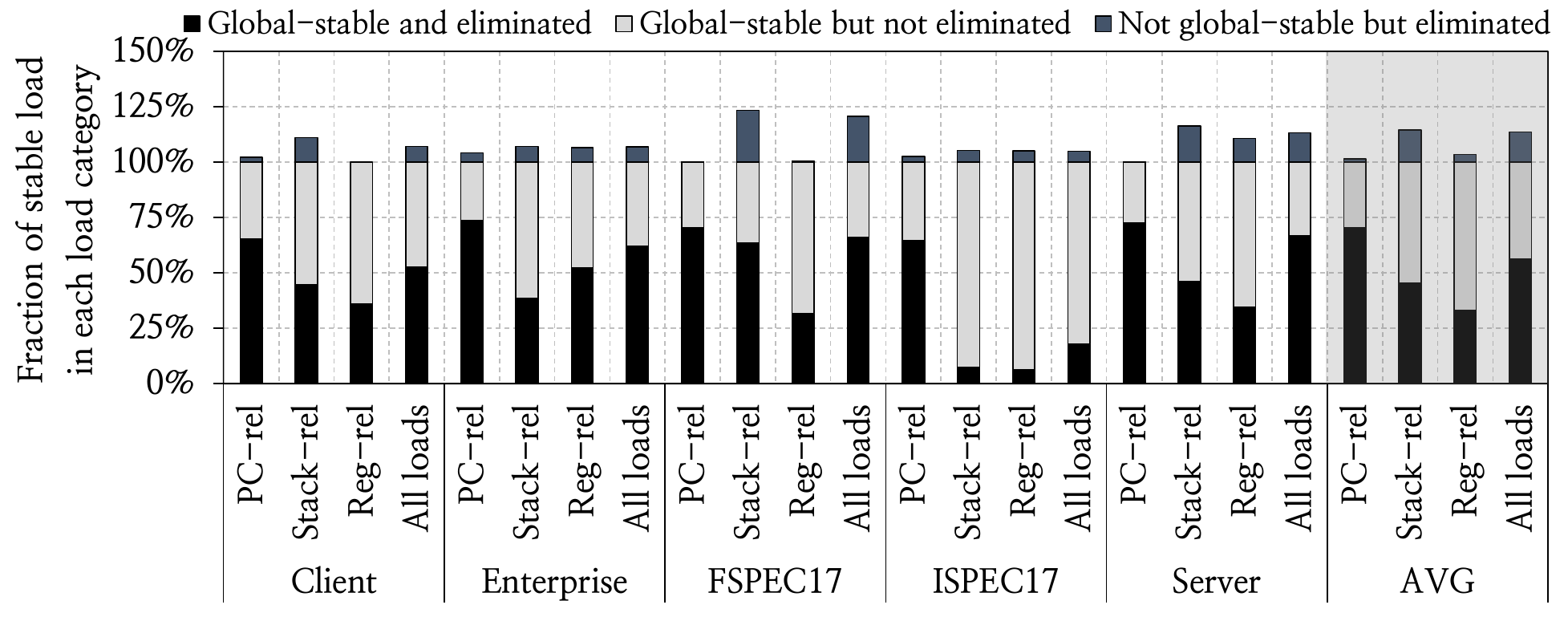}
\caption{
Breakdown of eliminated and non-eliminated loads as a fraction of \rbe{global-stable} loads.
}
\label{fig:cov_breakdown}
\end{figure}

To understand Constable's coverage of \rbe{global-stable} loads (see \mbox{\cref{sec:headroom}}), Fig.~\mbox{\ref{fig:cov_breakdown}} shows the breakdown of loads in each addressing-mode category into three classes: (1) loads that are \rbe{global-stable} and eliminated by Constable, (2) loads that are \rbe{global-stable} but not eliminated, and (3) loads that are not \rbe{global-stable} but eliminated.
We make three key observations.
First, PC-relative and register-relative \rbe{global-stable} loads see the highest and the lowest runtime elimination coverage of $70.2\%$ and $33.2\%$, respectively. 
Second, Constable successfully eliminates $56.4\%$ of all \rbe{global-stable} loads on average at runtime.
For the remaining $43.6\%$ \rbe{global-stable} loads, Constable misses their elimination opportunity due to three key reasons (not shown in the figure):
(a) at least one source architectural register of a \rbe{global-stable} load instruction gets written between its two successive dynamic instances (for $23.3\%$ of all \rbe{global-stable} loads), 
(b) a \emph{silent store}~\mbox{\cite{silent_stores,silent_stores2,silent_stores3,silent_stores4}} occurs between two successive dynamic instances of a \rbe{global-stable} load (for $14.1\%$ of all \rbe{global-stable} loads),
and (c) coverage loss due to other reasons, e.g., stability confidence learning and limited hardware budget for likely-stable load tracking (for $6.2\%$ of all \rbe{global-stable} loads).
Third, on average, $13.5\%$ more loads are eliminated by Constable at runtime which are not identified as \rbe{global-stable}. This is because these loads are not stable across the entire workload trace, but stable in a workload phase to meet the stability confidence threshold and and hence get eligible for elimination.
Based on these results, we conclude that Constable eliminates a significant fraction of the \rbe{global-stable} loads at runtime. 
However many elimination opportunities are still left, which can be unlocked by future works to achieve even higher performance and power efficiency improvements.

\subsection{Impact on Pipeline Resource Utilization} \label{sec:eval_pipe_resource_reduction}


\subsubsection{\textbf{Reduction in RS Allocation}} \label{sec:eval_reduction_rs}
\begin{figure}[!h] 
\centering
\includegraphics[width=3.4in]{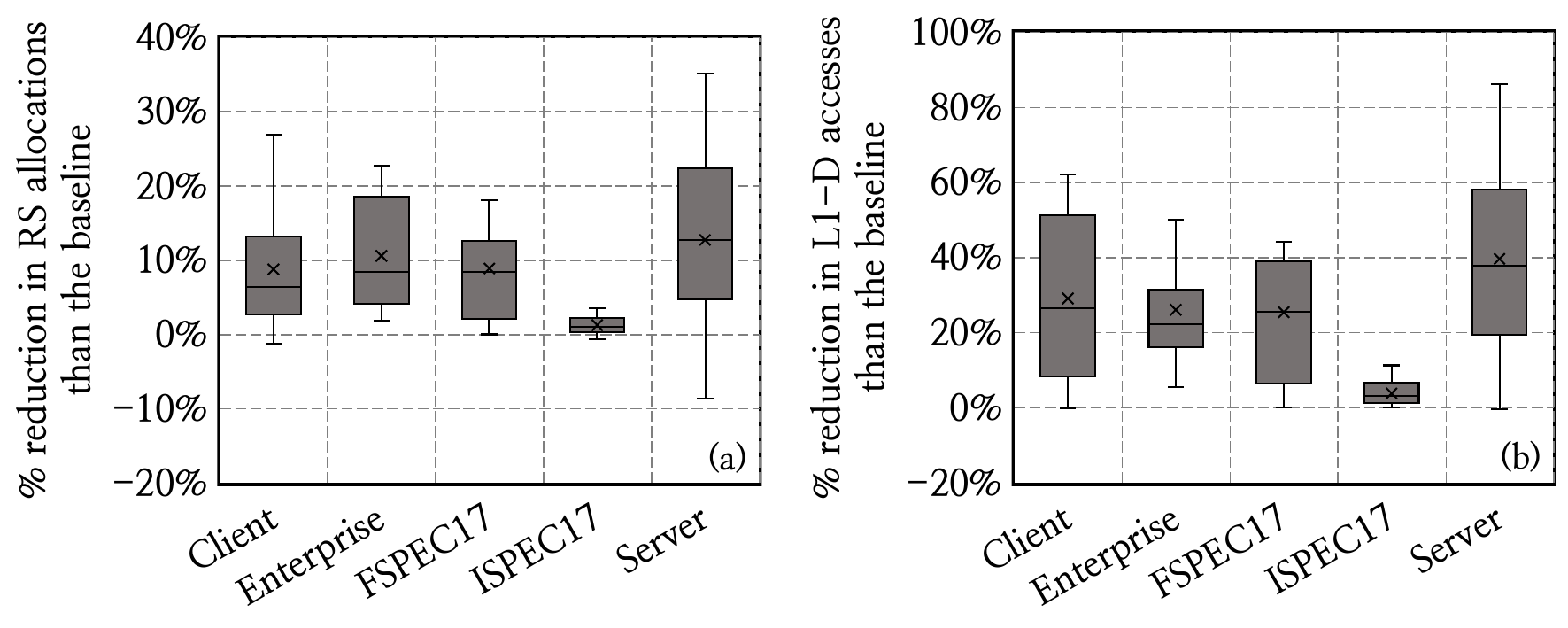}
\caption{Reduction in (a) RS allocations and (b) L1-D accesses. 
}
\label{fig:pipe_reduction}
\end{figure}

Fig.~\ref{fig:pipe_reduction}(a) plots the percentage reduction in RS allocations in a system with Constable over the baseline system as a box-and-whiskers plot. The key observation is that Constable reduces the RS allocation by $8.8\%$ on average (up to $35.1\%$) across all workloads. \texttt{Server} and \texttt{ISPEC17} workloads experience the highest and the lowest average RS allocation \rbd{reductions of} $12.8\%$ and $1.3\%$, respectively. 
\rbd{$37$ of the $90$ workloads experience a reduction in RS allocation by more than $10\%$}.

\subsubsection{\textbf{Reduction in L1-D Access}} \label{sec:eval_reduction_l1d}
Fig.~\ref{fig:pipe_reduction}(b) plots the percentage reduction in L1-D accesses in a system with Constable over the baseline as a box-and-whiskers plot. The key observation is that Constable reduces L1-D allocation by $26.0\%$ on average. Similar to the reduction in RS allocation, \texttt{Server} and \texttt{ISPEC17} workloads experience the highest and the lowest average L1-D access reduction of $39.7\%$ and $3.9\%$.

Based on these results, we conclude that, by eliminating load execution, Constable significantly reduces RS allocations and L1-D accesses, both of which aid in \rbd{improving performance (\cref{sec:eval_perf}) and} reducing dynamic power consumption (\cref{sec:eval_power}).  

\subsection{Power Improvement} \label{sec:eval_power}

\begin{figure}[!h] 
\centering
\includegraphics[width=3.4in]{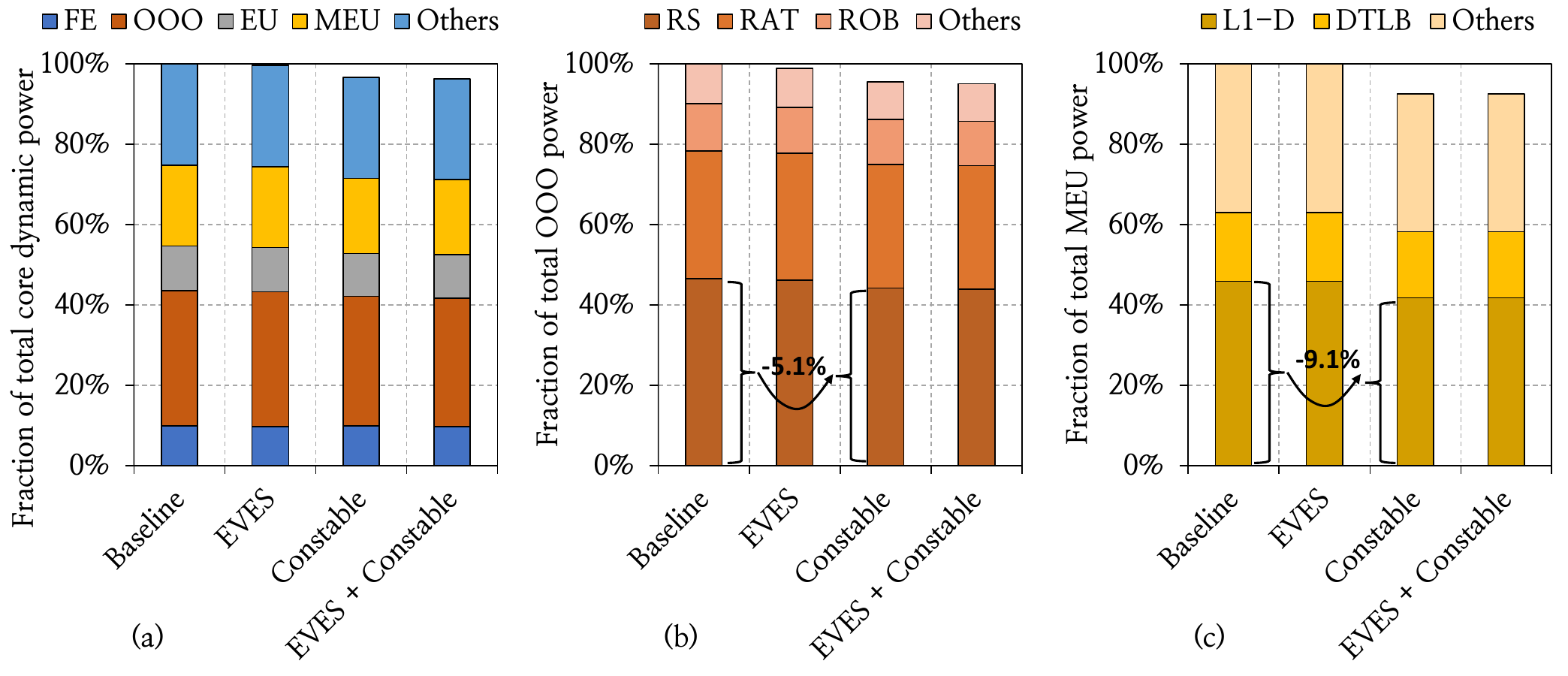}
\caption{(a) Overall core power consumption normalized to baseline. Expanded view of (b) OOO power and (c) MEU power.}
\label{fig:power}
\end{figure}

Fig.~\ref{fig:power}(a) shows the core power consumption \rbd{(and its breakdown) in} a system with EVES, Constable, and EVES+Constable normalized to the baseline. 
The key takeaway is that Constable reduces the core power consumption by $3.4\%$ on average over the baseline, whereas EVES reduces power by \rbd{only} $0.2\%$. 
This is because, unlike EVES where the value-predicted load instructions get executed nonetheless, Constable eliminates executing likely-stable loads altogether. 
To understand the distribution of the power benefit across various core structures, we further expand the power consumption \rbd{of} OOO and MEU units in Fig.~\ref{fig:power}(b) and (c) respectively. 
As Fig.~\ref{fig:power}(b) shows, Constable reduces the power consumed by OOO unit by $4.5\%$ on average over the baseline. 
The RS sub-unit of OOO unit experiences the highest power reduction of $5.1\%$ (marked by braces). 
This is because Constable significantly reduces the number of RS allocations (\cref{sec:eval_reduction_rs}).
As Fig.~\ref{fig:power}(c) shows, Constable also reduces the power consumed by MEU unit by $7.2\%$ on average over the baseline. 
The MEU power reduction is dominated by L1-D cache, which experiences $9.1\%$ reduction in power (marked by braces) on average. This is largely due to the reduction is L1-D accesses (\cref{sec:eval_reduction_l1d}).
\rbd{We} conclude that Constable, unlike value prediction, reduces the core power by fundamentally eliminating load execution.

\section{Other Related Works} \label{sec:related_works}

\vspace{2pt}
\noindent \textbf{Compiler-based optimizations} like global value numbering~\cite{rosen1988global,click1995global}, common subexpression elimination~\cite{cocke1970global}, and loop-invariant code motion~\cite{hoe1986compilers} identify pieces of code that stay invariant across various \rbd{granularities} of a program 
and eliminate their redundant execution. 
All the workloads used in this work are already compiled with these optimizations. 
We demonstrate that Constable provides performance benefits on top of such well-optimized workloads.

\vspace{2pt}
\noindent \textbf{Data caching}~\cite{wilkes1965slave,liptay1968structural,cache_ibm} reduces the average memory access latency by storing the data that would likely get accessed in the near future in faster on-chip memory. 
Prior works have proposed techniques to improve cache hit-rate by (1) exploiting data reuse patterns (e.g.,~\cite{qureshi2007adaptive,jaleel2010high,pacman,pa_haweye,hawkeye,catch,mockingjay}), (2) predicting dead cachelines (e.g.,~\cite{sdbp,jimenez2017multiperspective,teran2016perceptron}), and (3) applying machine learning techniques(e.g.,~\cite{glider,parrot,chrome}).
Constable can be orthogonally combined with caching techniques. \rbd{Our baseline system employs a dead-block-aware replacement policy in the LLC (Table~\ref{table:sim_params})}.

\vspace{2pt}
\noindent A \textbf{data prefetcher} predicts the address of future memory requests and fetches the data from slower memory to faster on-chip caches before the program demands it. A prefetch request can be generated by software~\cite{software_pref,software_preexecute,ainsworth2017software,ainsworth2016graph,rpg2} or hardware. Hardware prefetchers can generate prefetch request by (1) pre-executing program code~\cite{dundas,mutlu2003runahead,mutlu2003runahead2, mutlu2005techniques,mutlu2006efficient,hashemi2016continuous,mutlu2005address,hashemi2015filtered,vector_runahead,mutlu2005reusing,iacobovici2004effective,precise_runahead,decoupled_vector_runahead}, (2) learning memory access pattern over spatial memory regions~\cite{stride,streamer,baer2,jouppi_prefetch,ampm,fdp,footprint,sms,spp,vldp,sandbox,bop,dol,dspatch,bingo,mlop,ppf,litz2022crisp,pmp,pythia,navarro2022berti,clip,pathfinder}, and (3) memorizing long sequences of past demanded memory addresses~\cite{markov,stems,somogyi_stems,wenisch2010making,domino,isb,misb,triage,wenisch2005temporal,chilimbi2002dynamic,chou2007low,ferdman2007last,hu2003tcp,bekerman1999correlated,cooksey2002stateless,karlsson2000prefetching,voyager}.
Constable can be orthogonally applied with any prefetcher.
\rbd{Our baseline system employs multiple data prefetchers across the cache hierarchy (Table~\ref{table:sim_params}).}

\section{Conclusion}

We introduce Constable, a purely-microarchitectural technique that safely eliminates the execution of load instructions while breaking the load data dependence.
Our extensive \rbd{evaluation} using a wide range of workloads and system configurations shows that Constable provides significant performance benefit \rbd{and reduced} dynamic power consumption by eliminating load execution.
As hardware resource scaling becomes challenging in future processors, we believe and hope that Constable's key observations and insights would inspire future works to explore a multitude of other optimizations that mitigates ILP loss due to resource dependence \rbd{and load instruction execution}.

\section*{Acknowledgments}
\ifarxiv
\rbf{This work was partially done when Rahul Bera was an intern at Intel Processor Architecture Research Lab. Concepts, techniques, and implementations presented in this paper may be subject matter of pending patent applications filed by Intel Corporation.}
\fi
We thank Rakesh Kumar (NTNU) and anonymous referees of ISCA 2024 for their valuable insights and feedback on this work.
We thank the SAFARI Research Group members for providing a stimulating intellectual environment.
\rbf{This work is supported in part by Semiconductor Research Corporation.}
Rahul thanks his departed father, whom he misses dearly everyday.

\newpage

\bibliographystyle{IEEEtranS}
\bibliography{refs}

\clearpage
\appendix
\section{Extended Evaluation} \label{sec:arxiv_ext_eval}

\subsection{Performance Sensitivity} \label{sec:arxiv_ext_eval_sensitivity}

\subsubsection{\textbf{Sensitivity to Load Execution Width Scaling}} \label{sec:arxiv_ext_eval_sen_width}

\begin{figure}[!h] 
\centering
\includegraphics[width=3.4in]{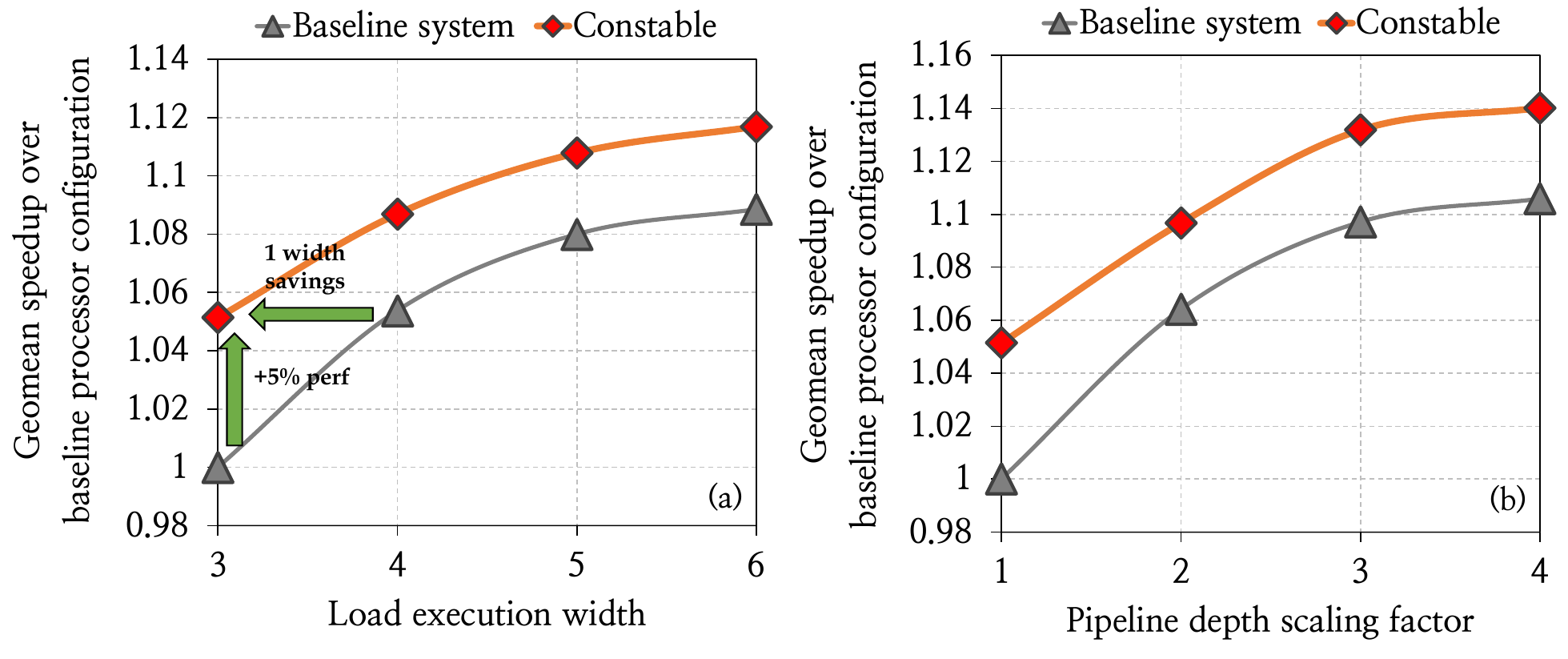}
\caption{Performance sensitivity to (a) load execution width, and (b) pipeline depth.}
\label{fig:sensitivity}
\end{figure}

Fig.~\ref{fig:sensitivity}(a) shows the geomean speedup of Constable and the baseline system over the baseline configuration when we increase the load execution width (i.e., increasing both number of AGU and load ports). 
We make two key observations.
First, Constable \emph{consistently} adds performance on top of the baseline system even if we naively scale the load execution width. 
With increasing AGU and load ports (while keeping the pipeline depth resources same), the resource dependence stemming from load reduces. Yet, Constable outperforms the baseline system by $3.5\%$ with $2\times$ load execution width than the baseline configuration.
Second, adding Constable on the baseline system configuration (i.e., with 3 load execution width) essentially provides the similar performance benefit as the baseline system with one extra load execution width, while incurring lower area overhead and \emph{reducing} power consumption.

\subsubsection{\textbf{Sensitivity to Pipeline Depth Scaling}} \label{sec:arxiv_ext_eval_sen_depth}
Fig.~\ref{fig:sensitivity}(a) shows the geomean speedup of Constable and the baseline system over the baseline configuration when we scale pipeline depth resources (i.e., size of ROB, RS, LB and SB).
The key takeaway is that Constable \emph{consistently} adds performance on top of the baseline system even if we naively scale the pipeline depth. With $4\times$ depth scaling, Constable improves performance of the baseline system by $3.4\%$ on average.

\subsection{Effect of Eliminated Load Disambiguation \\ with In-Flight Stores} \label{sec:arxiv_ext_eval_pipe_flush}

\begin{figure}[!h] 
\centering
\includegraphics[width=3.4in]{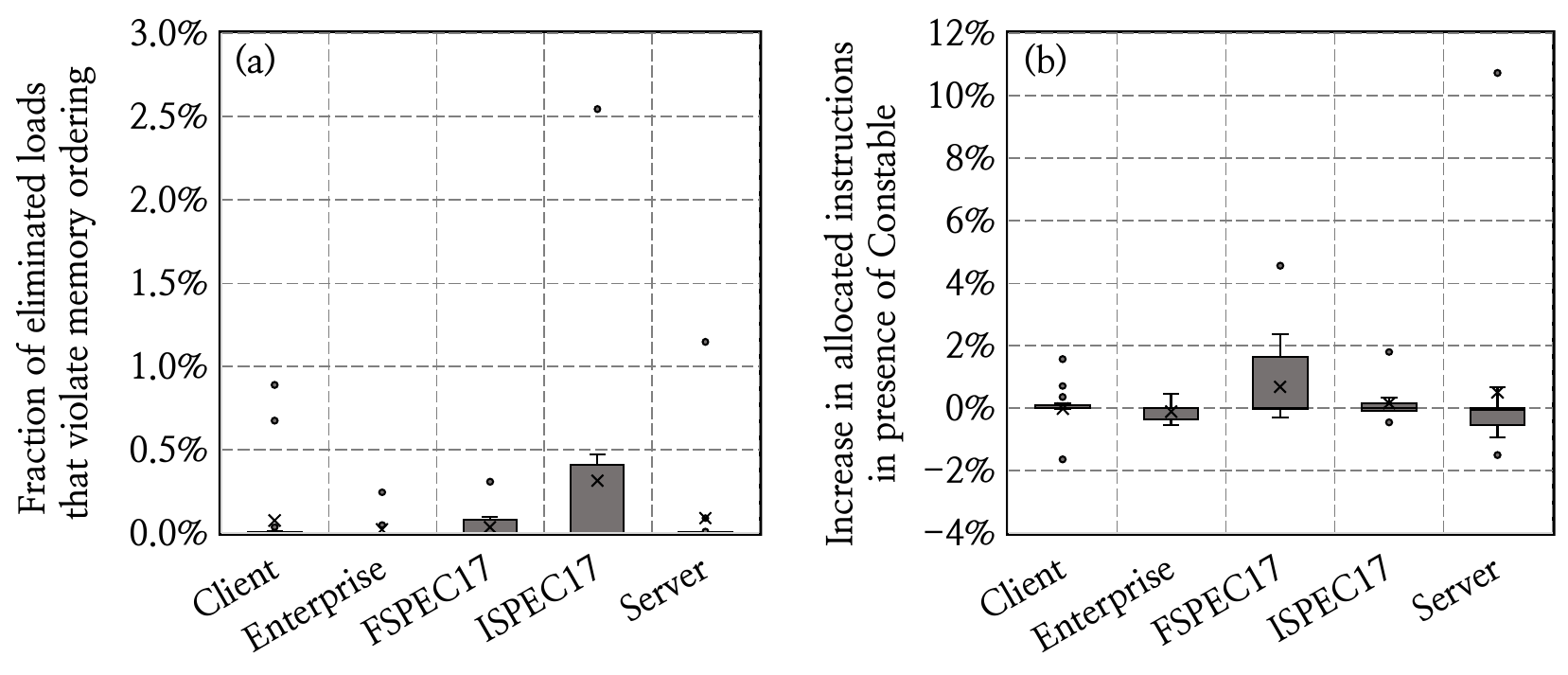}
\caption{(a) Fraction of loads eliminated by Constable that violate memory ordering. (2) Increase in instructions allocated to ROB in presence of Constable.}
\label{fig:elim_nuke}
\end{figure}

When the computed address of an in-flight store instruction matches with that of an eliminated load younger than the store, the existing memory disambiguation logic catches such memory ordering violation and re-executes all instructions younger than (and including) the incorrectly-eliminated load (see~\cref{sec:design_in_flight_stores}). 
Thus a frequent memory ordering violation by eliminated loads may incur a significant performance and power overhead on Constable.
To understand such overhead, we show the fraction of loads eliminated by Constable that violate memory ordering as a box-and-whiskers plot in Fig.~\ref{fig:elim_nuke}(a). 
As we can see, an eliminated load rarely violates memory ordering. On average, only $0.09\%$ of all eliminated loads violate memory ordering. Less than $0.5\%$ of eliminated loads violate memory ordering in $86$ out of $90$ workloads.
This is primarily due to Constable's confidence-based mechanism that considers a load instruction eligible for elimination only if it meets a sufficiently-high stability confidence level threshold (set to 30 in our evaluation).
Fig.~\ref{fig:elim_nuke}(b) further shows the increase in instructions allocated to the ROB in presence of Constable as compared to the baseline system to understand the effect of such rare memory ordering violations. As we can see, Constable increases the allocated instructions by only $0.3\%$ on average across all workloads. $79$ out of $90$ workloads observe an increase of less than $1\%$.

Thus, we conclude that Constable observes a very insignificant overhead due to rare memory ordering violations by incorrectly-eliminated loads.

\subsection{Effect of Loss of Elimination Opportunities \\ due to Clean Evictions} \label{sec:arxiv_eval_clean_eviction}

In order to correctly eliminate load instructions in a multi-core system, Constable proposes pinning the CV-bit of a cacheline that is accessed by an eliminated load instruction (see~\cref{sec:design_multi_core_coherence}). 
However, the change in the coherence protocol may complicate hardware verification.
Another alternative design could be to avoid elimination on every core-private cache eviction. However, this design choice may lose elimination opportunities if the evicted cacheline is clean. In this section, we quantify impact of such elimination opportunity loss on Constable's performance and elimination coverage.

\begin{figure}[!h] 
\centering
\includegraphics[width=3.4in]{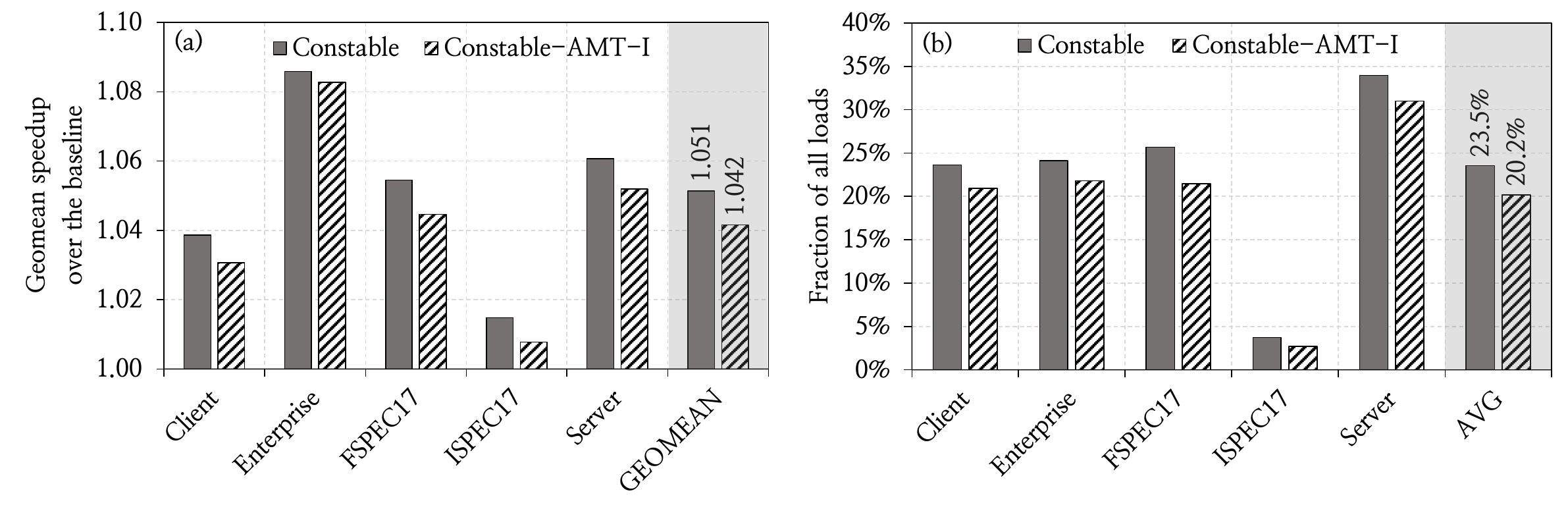}
\caption{(a) Speedup and (b) coverage of Constable with AMT invalidation on L1D eviction compared to a vanilla Constable.}
\label{fig:clean_eviction_effect}
\end{figure}

To understand the effect, we model a Constable variant that looks up AMT for every L1 data (L1-D) cache eviction and invalidates the AMT entry. This prevents Constable from eliminating any further load instructions that access the evicted cacheline. We call this Constable variant Constable-AMT-I.
Fig.~\ref{fig:clean_eviction_effect}(a) compares the speedup of Constable-AMT-I with the vanilla Constable. 
Constable-AMT-I loses $0.9\%$ performance improvement than the vanilla Constable on average across all workloads. $11$ out of $90$ workloads observe a performance loss of more than $5\%$ in Constable-AMT-I (with the highest performance loss of $10.4\%$ in \texttt{554.roms\_r}) as compared to vanilla Constable.
The performance loss is primarily attributed to the loss in elimination coverage. As Fig.~\ref{fig:clean_eviction_effect}(b) shows, Constable-AMT-I provides $3.4\%$ less load elimination coverage than the vanilla Constable. $17$ out of $90$ workloads observe a coverage loss of more than $5\%$ in Constable-AMT-I (with the highest coverage loss of $27\%$ in \texttt{554.roms\_r}).
Thus, we conclude that CV-bit pinning is a more-performant design choice than avoiding elimination on every core-private cache eviction.

\section{Effect of Increasing Architectural Registers on \rbe{Global-Stable} Loads} \label{sec:arxiv_incr_arch_reg}

Increasing architectural registers enables a compiler to exploit additional registers to capture data reuse, which otherwise would have been reused via memory. Thus increasing architectural registers typically reduces the number of load and store instructions in a program.
To understand the effect of increasing architectural registers on \rbe{global-stable} load instructions, we compile all C/C++-based workloads from SPEC CPU 2017 rate suite~\cite{spec2017} without and with Intel APX extension~\cite{intel_apx}, that doubles the number of architectural registers in x86-64 ISA from 16 to 32, using Clang 18.1.3~\cite{clang_18}.
We run these workloads using the test input and profile their \emph{end-to-end execution} using the Load Inspector tool (see \cref{sec:headroom_stable_loads_why}) to observe (1) the reduction in dynamic loads caused by APX, and (2) the fraction of all dynamic loads that are \rbe{global-stable} in workloads without and with APX.

\begin{figure}[!h] 
\centering
\includegraphics[width=3.4in]{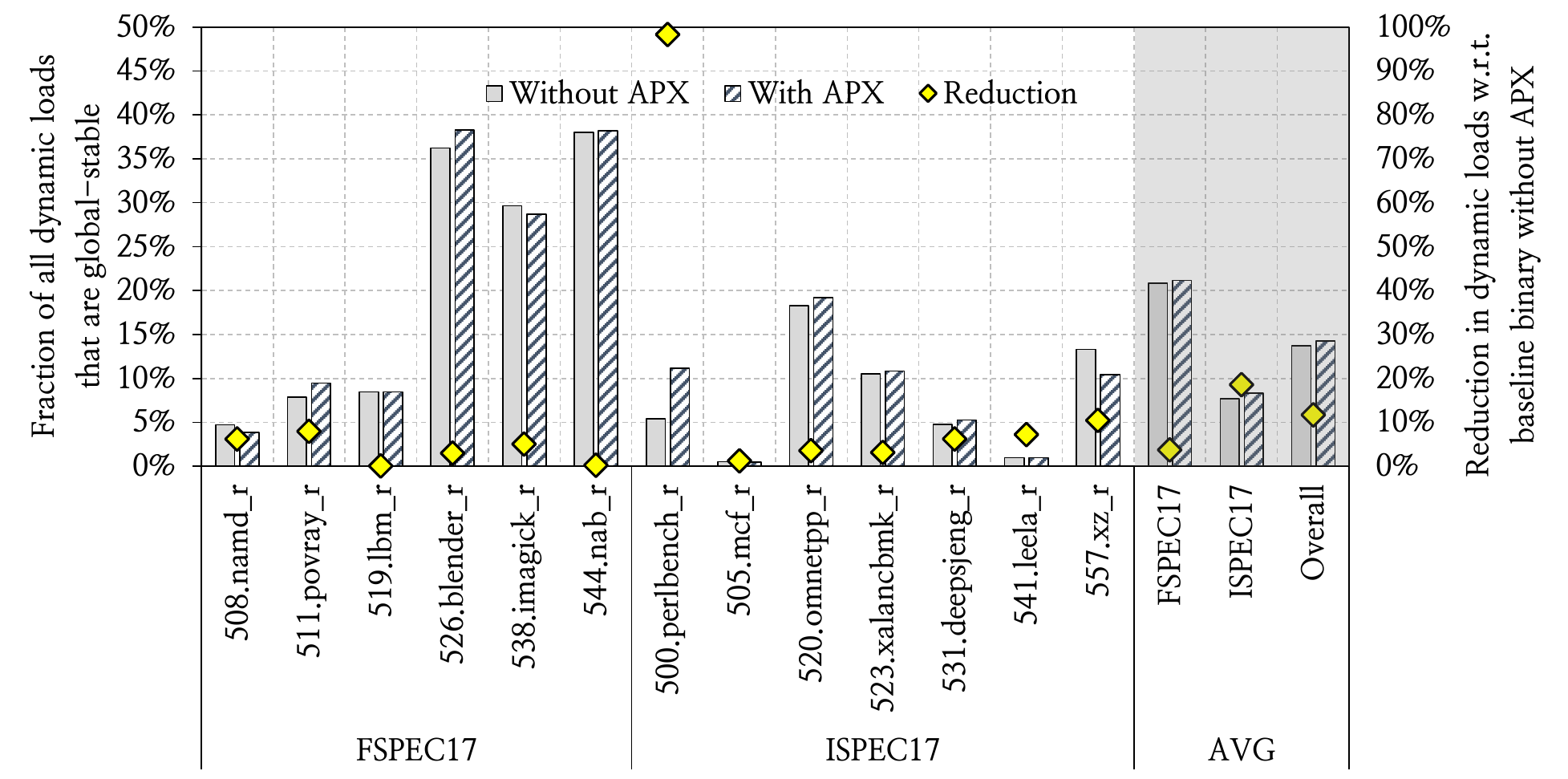}
\caption{Fraction of all dynamic loads that are \rbe{global-stable} in workloads compiled without and with APX (on the left y-axis) and the reduction in dynamic loads by APX (on the right y-axis).}
\label{fig:apx_no_apx}
\end{figure}

Fig.~\ref{fig:apx_no_apx} shows the fraction of all dynamic loads that are \rbe{global-stable} in each workload without and with APX (as bar graph on the left y-axis) and the reduction in dynamic loads with APX extension (as markers on the right y-axis) for all C/C++-based workloads from SPEC CPU 2017 suite.\footnote{\texttt{502.gcc\_r}, \texttt{510.parest\_r} and \texttt{525.x264\_r} are omitted from this study due to failed compilation using Clang.}

We make two key observations from Fig.~\ref{fig:apx_no_apx}.
First, the fraction of dynamic loads that are \rbe{global-stable} (i.e., the elimination opportunities for Constable) is much higher than the reduction in dynamic loads by doubling the number of architectural registers. APX reduces the number of dynamic loads by $11.7\%$ on average. \texttt{500.perlbench\_r} is an outlier that observes a reduction of $98.3\%$ of loads. Without it, APX reduces the dynamic loads by only $4.5\%$ on average. On the other hand, $13.7\%$ and $14.2\%$ of all dynamic loads on average are \rbe{global-stable} in workloads without and with APX, respectively.\footnote{Note that, the \rbe{global-stable} load fraction reported here is slightly lower than that reported in Fig.~\ref{fig:stable_load_characterization}(a). This is because, the earlier study in Fig.~\ref{fig:stable_load_characterization}(a) uses the representative sections of the workloads (see~\cref{sec:methodology_workloads}) to limit the simulation overhead, while this study instruments each workload \emph{end-to-end}.}
The difference is more prominent for \texttt{FSPEC17} workloads, where APX reduces dynamic loads by only $3.7\%$, whereas $20.8\%$ and $21.2\%$ of dynamic loads are \rbe{global-stable} in without and with APX, respectively.
Second, the fraction of dynamic loads that are \rbe{global-stable} is nearly the same in workloads without and with APX. \texttt{500.perlbench\_r} and \texttt{557.xz\_r} are only two workloads that show more than $3\%$ absolute change in the \rbe{global-stable} load fraction.
This shows that the elimination opportunities for Constable is largely orthogonal to the benefits of increasing architectural registers.

\begin{figure}[!h] 
\centering
\includegraphics[width=3.4in]{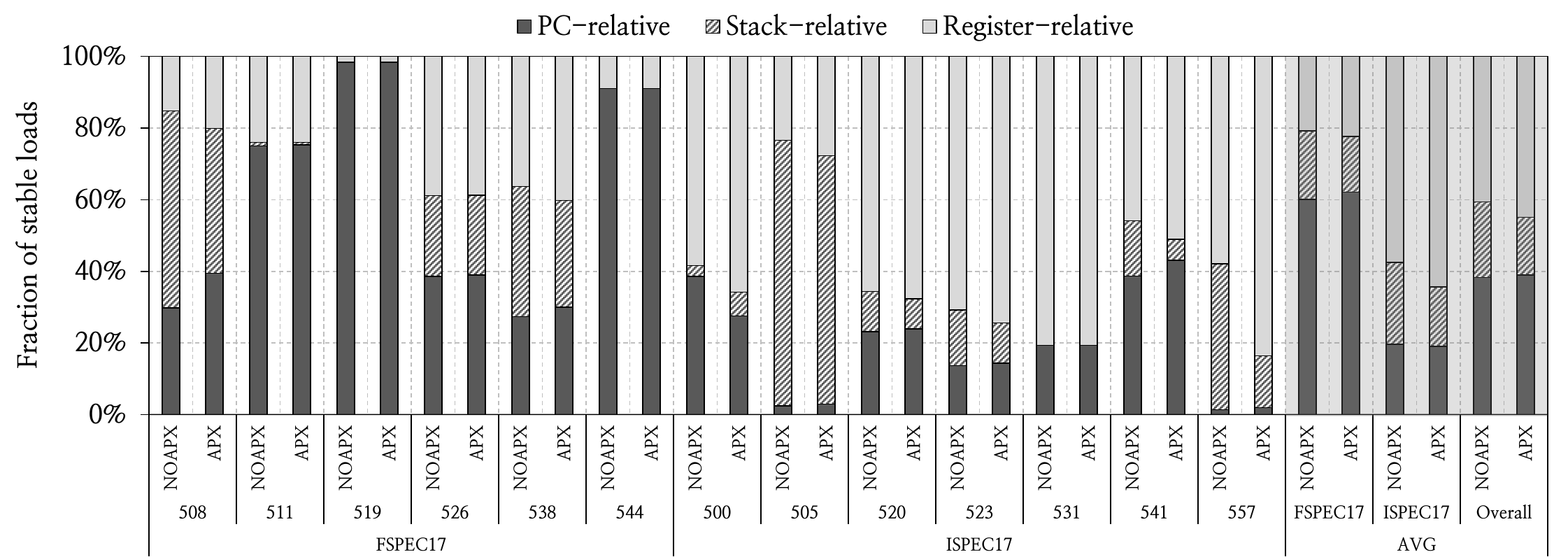}
\caption{Distribution of \rbe{global-stable} loads by their addressing modes in workloads without and with APX. Each number on the x-axis corresponds to the respective workload from SPEC CPU 2017 suite.}
\label{fig:apx_no_apx_addr_mode}
\end{figure}

To further analyze the change in characteristics of \rbe{global-stable} loads in presence of APX, we break down the \rbe{global-stable} loads based on their addressing modes in workloads both without and with APX in Fig.~\ref{fig:apx_no_apx_addr_mode}.
We make two key observations from this figure.
First, the fraction of stack-relative \rbe{global-stable} loads reduces in presence of APX. On average, $21.1\%$ and $16\%$ of all \rbe{global-stable} loads use stack-relative addressing in workloads without and with APX, respectively. This is expected, since increasing architectural registers predominantly reduces stack loads.
Second, the fraction of PC-relative \rbe{global-stable} loads stays nearly the same in presence of APX ($38.3\%$ without APX as compared to $38.9\%$ with APX). This shows that doubling architectural registers alone cannot eliminate all memory accesses to global-scope variables which are effectively runtime constant.

Based on these results, we conclude that the two load elimination techniques - at compile time by increasing architectural registers and at runtime by Constable - are largely \emph{orthogonal} to each other. Thus, Constable would likely be equally-performant and power-efficient in presence of increased architectural registers, as it is with the current set of architectural registers.

\end{document}